\documentclass[twocolumn, article, times, 12pt]{aastex631}
\usepackage[utf8]{inputenc}
\usepackage{graphicx}
\usepackage{mathtools}
\usepackage{natbib}  
\usepackage{comment}
\usepackage{float}
\usepackage{hyperref}
\usepackage{booktabs}
\newcommand{\figures}{.\figures}

\expandafter\def\expandafter\normalsize\expandafter{%
    \normalsize%
\setlength\abovedisplayskip{1pt}%
\setlength\belowdisplayskip{1pt}%
\setlength\abovedisplayshortskip{1pt}%
\setlength\belowdisplayshortskip{1pt}%
}

\newcommand{\styfileloc}{styfiles}

\usepackage{\styfileloc/j1903scatt}

\begin{document}
\title{The NANOGrav 12.5-Year Data Set: Probing Interstellar Turbulence and Precision Pulsar Timing with PSR J1903+0327}
\author[0009-0004-0992-9040]{Abra Geiger}
\affiliation{Cornell Center for Astrophysics and Planetary Science and Department of Astronomy, Cornell University, Ithaca, NY 14853, USA}
\author[0000-0002-4049-1882]{James M. Cordes}
\affiliation{Cornell Center for Astrophysics and Planetary Science and Department of Astronomy, Cornell University, Ithaca, NY 14853, USA}
\author[0000-0003-0721-651X]{Michael T. Lam}
\affiliation{SETI Institute, 339 N Bernardo Ave Suite 200, Mountain View, CA 94043, USA}
\affiliation{School of Physics and Astronomy, Rochester Institute of Technology, Rochester, NY 14623, USA}
\affiliation{Laboratory for Multiwavelength Astrophysics, Rochester Institute of Technology, Rochester, NY 14623, USA}
\author[0000-0002-4941-5333]{Stella Koch Ocker}
\affiliation{Division of Physics, Mathematics, and Astronomy, California Institute of Technology, Pasadena, CA 91125, USA}
\affiliation{The Observatories of the Carnegie Institution for Science, Pasadena, CA 91101, USA}
\author[0000-0002-2878-1502]{Shami Chatterjee}
\affiliation{Cornell Center for Astrophysics and Planetary Science and Department of Astronomy, Cornell University, Ithaca, NY 14853, USA}
\author{Zaven Arzoumanian}
\affiliation{X-Ray Astrophysics Laboratory, NASA Goddard Space Flight Center, Code 662, Greenbelt, MD 20771, USA}
\author[0009-0008-9349-1086]{Ava L. Battaglia}
\affiliation{Department of Physics, Lafayette College, Easton, PA 18042, USA}
\author[0000-0003-4046-884X]{Harsha Blumer}
\affiliation{Department of Physics and Astronomy, West Virginia University, P.O. Box 6315, Morgantown, WV 26506, USA}
\affiliation{Center for Gravitational Waves and Cosmology, West Virginia University, Chestnut Ridge Research Building, Morgantown, WV 26505, USA}
\author[0000-0003-3053-6538]{Paul R. Brook}
\affiliation{Institute for Gravitational Wave Astronomy and School of Physics and Astronomy, University of Birmingham, Edgbaston, Birmingham B15 2TT, UK}
\author[0009-0002-1861-9787]{Olivia A. Combs}
\affiliation{Department of Physics, Lafayette College, Easton, PA 18042, USA}
\author[0000-0002-6039-692X]{H. Thankful Cromartie}
\affiliation{National Research Council Research Associate, National Academy of Sciences, Washington, DC 20001, USA resident at Naval Research Laboratory, Washington, DC 20375, USA}
\author[0000-0002-2185-1790]{Megan E. DeCesar}
\affiliation{George Mason University, Fairfax, VA 22030, resident at the U.S. Naval Research Laboratory, Washington, DC 20375, USA}
\author[0000-0002-6664-965X]{Paul B. Demorest}
\affiliation{National Radio Astronomy Observatory, 1003 Lopezville Rd., Socorro, NM 87801, USA}
\author[0000-0001-8885-6388]{Timothy Dolch}
\affiliation{Department of Physics, Hillsdale College, 33 E. College Street, Hillsdale, MI 49242, USA}
\affiliation{Eureka Scientific, 2452 Delmer Street, Suite 100, Oakland, CA 94602-3017, USA}
\author{Justin A. Ellis}
\altaffiliation{Infinia ML, 202 Rigsbee Avenue, Durham NC, 27701, USA}
\author[0000-0002-2223-1235]{Robert D. Ferdman}
\affiliation{School of Chemistry, University of East Anglia, Norwich, NR4 7TJ, United Kingdom}
\author[0000-0001-7828-7708]{Elizabeth C. Ferrara}
\affiliation{Department of Astronomy, University of Maryland, College Park, MD 20742, USA}
\affiliation{Center for Research and Exploration in Space Science and Technology, NASA/GSFC, Greenbelt, MD 20771}
\affiliation{NASA Goddard Space Flight Center, Greenbelt, MD 20771, USA}
\author[0000-0001-8384-5049]{Emmanuel Fonseca}
\affiliation{Department of Physics and Astronomy, West Virginia University, P.O. Box 6315, Morgantown, WV 26506, USA}
\affiliation{Center for Gravitational Waves and Cosmology, West Virginia University, Chestnut Ridge Research Building, Morgantown, WV 26505, USA}
\author[0000-0001-6166-9646]{Nate Garver-Daniels}
\affiliation{Department of Physics and Astronomy, West Virginia University, P.O. Box 6315, Morgantown, WV 26506, USA}
\affiliation{Center for Gravitational Waves and Cosmology, West Virginia University, Chestnut Ridge Research Building, Morgantown, WV 26505, USA}
\author[0000-0001-8158-683X]{Peter A. Gentile}
\affiliation{Department of Physics and Astronomy, West Virginia University, P.O. Box 6315, Morgantown, WV 26506, USA}
\affiliation{Center for Gravitational Waves and Cosmology, West Virginia University, Chestnut Ridge Research Building, Morgantown, WV 26505, USA}
\author[0000-0003-1884-348X]{Deborah C. Good}
\affiliation{Department of Physics and Astronomy, University of Montana, 32 Campus Drive, Missoula, MT 59812}
\author[0000-0001-6607-3710]{Megan L. Jones}
\affiliation{Center for Gravitation, Cosmology and Astrophysics, Department of Physics, University of Wisconsin-Milwaukee,\\ P.O. Box 413, Milwaukee, WI 53201, USA}
\author[0000-0003-1301-966X]{Duncan R. Lorimer}
\affiliation{Department of Physics and Astronomy, West Virginia University, P.O. Box 6315, Morgantown, WV 26506, USA}
\affiliation{Center for Gravitational Waves and Cosmology, West Virginia University, Chestnut Ridge Research Building, Morgantown, WV 26505, USA}
\author[0000-0001-5373-5914]{Jing Luo}
\altaffiliation{Deceased}
\affiliation{Department of Astronomy \& Astrophysics, University of Toronto, 50 Saint George Street, Toronto, ON M5S 3H4, Canada}
\author[0000-0001-5229-7430]{Ryan S. Lynch}
\affiliation{Green Bank Observatory, P.O. Box 2, Green Bank, WV 24944, USA}
\author[0000-0001-7697-7422]{Maura A. McLaughlin}
\affiliation{Department of Physics and Astronomy, West Virginia University, P.O. Box 6315, Morgantown, WV 26506, USA}
\affiliation{Center for Gravitational Waves and Cosmology, West Virginia University, Chestnut Ridge Research Building, Morgantown, WV 26505, USA}
\author[0000-0002-3616-5160]{Cherry Ng}
\affiliation{Dunlap Institute for Astronomy and Astrophysics, University of Toronto, 50 St. George St., Toronto, ON M5S 3H4, Canada}
\author[0000-0002-6709-2566]{David J. Nice}
\affiliation{Department of Physics, Lafayette College, Easton, PA 18042, USA}
\author[0000-0001-5465-2889]{Timothy T. Pennucci}
\affiliation{Institute of Physics and Astronomy, E\"{o}tv\"{o}s Lor\'{a}nd University, P\'{a}zm\'{a}ny P. s. 1/A, 1117 Budapest, Hungary}
\author[0000-0002-8826-1285]{Nihan S. Pol}
\affiliation{Department of Physics, Texas Tech University, Box 41051, Lubbock, TX 79409, USA}
\author[0000-0001-5799-9714]{Scott M. Ransom}
\affiliation{National Radio Astronomy Observatory, 520 Edgemont Road, Charlottesville, VA 22903, USA}
\author[0000-0002-6730-3298]{Ren\'{e}e Spiewak}
\affiliation{Jodrell Bank Centre for Astrophysics, University of Manchester, Manchester, M13 9PL, United Kingdom}
\author[0000-0001-9784-8670]{Ingrid H. Stairs}
\affiliation{Department of Physics and Astronomy, University of British Columbia, 6224 Agricultural Road, Vancouver, BC V6T 1Z1, Canada}
\author[0000-0002-7261-594X]{Kevin Stovall}
\affiliation{National Radio Astronomy Observatory, 1003 Lopezville Rd., Socorro, NM 87801, USA}
\author[0000-0002-1075-3837]{Joseph K. Swiggum}
\altaffiliation{NANOGrav Physics Frontiers Center Postdoctoral Fellow}
\affiliation{Department of Physics, Lafayette College, Easton, PA 18042, USA}
\author[0000-0003-4700-9072]{Sarah J. Vigeland}
\affiliation{Center for Gravitation, Cosmology and Astrophysics, Department of Physics, University of Wisconsin-Milwaukee,\\ P.O. Box 413, Milwaukee, WI 53201, USA}

\begin{abstract}
\noindent Free electrons in the interstellar medium refract and diffract radio waves along multiple paths, resulting in  angular and temporal broadening of radio pulses that limits pulsar timing precision.  
We determine multifrequency, multi-epoch scattering times for the large dispersion measure millisecond pulsar J1903+0327 by developing a three component model for the emitted pulse shape that is convolved with a best fit pulse broadening function (PBF) identified from a family of  thin-screen and extended-media PBFs. We show that the scattering time, $\tau$, at a fiducial frequency of 1500 MHz changes by approximately 10\% over a 5.5\,yr span with a characteristic timescale of approximately 100 days. We also constrain the spectral index and inner scale of the wavenumber spectrum of electron density variations along this line of sight. We find that the scaling law for $\tau$ vs. radio frequency is strongly affected by any mismatch between the true and assumed PBF or between the true and assumed intrinsic pulse shape. We show using simulations that refraction is a plausible cause of the epoch dependence of $\tau$, manifesting as changes in the PBF shape and $1/e$ time scale. Finally, we discuss the implications of our scattering results on pulsar timing including time of arrival delays and dispersion measure misestimation.
\end{abstract}

\section{Introduction}
\label{sec:intro}

The spin stability of millisecond pulsars (MSPs) enables their use as astrophysical clocks in pulsar timing programs that probe plasma propagation processes and gravity.   The most notable application is the recently announced evidence for a stochastic background of nanohertz gravitational waves (GWs), most likely produced by a population of supermassive binary black holes at cosmological distances \citep{2023ApJ...951L...8A,EPTAGW23,2023ApJ...951L...6R}.

The timing program conducted by NANOGrav (the North American Nanohertz Observatory for Gravitational Waves) yields pulse times of arrival with sub-microsecond precision for many of the 68 pulsars reported in \citet{2023ApJ...951L...8A}.  Forward goals of the program include better characterization of the GW spectrum at yet lower frequencies and the detection of individual supermassive black hole binary systems.  These require longer data sets at monthly cadence or faster, combined with arrival times measured with higher precision.     

Arrival time  precision involves a combination of  astrophysical and instrumental factors,
the latter including the growing problem of radio frequency interference \citep{2023ApJ...951L..10A}. MSPs can be selected on the basis of their spin stabilities and the best objects are those with the narrowest pulses and shortest periods, yielding  a large number of pulses per unit observing time.  Propagation through the interstellar  medium (ISM) yields several kinds of chromatic delays that need to be mitigated in order to improve timing precision \citep{rickett_ism_prop, 2023ApJ...951L..10A}.  
More robust techniques for correction of the impact of interstellar scattering on the arrival times reported by pulsar timing array (PTA) groups are warranted. 
In this paper we use simulations and analysis of a highly scattered millisecond pulsar to study pulse shape distortions  produced by  multipath propagation through the ISM and  the resulting arrival time biases that result.  

The largest delays from the ISM are dispersion delays 
$\propto \nu^{-2}\DM$ that have a (nearly) deterministic dependence on 
radio frequency $\nu$, where dispersion measure, $\DM$, is the line of sight (\los) integral of the electron density, $\nelec$.  Removal of dispersion delays requires a contemporaneous measurement of DM because it is time variable due to spatial variations of the electron density
encountered along the \los. 
Inhomogeneities in the  electron number density diffract and refract pulses (which we collectively describe as `scattering.').  This causes broadening and angular wandering  of a pulsar's image, pulse distortion, and an overall delay that scales approximately as $\nu^{-4}$.  Estimation of \DM\ using multifrequency measurements is complicated by the presence of scattering delays, which nontrivially evolve across frequency in concert with the emitted pulse shape.
In particular, we show that the scattering delay contribution to an arrival time differs not only from the amplitude of the scattering time but also in its scaling with frequency.   This means that using multifrequency arrival times to estimate scattering times will often lead to misestimated scattering times.

Our analysis focuses on the PSR J1903+0327 ($P= 2.15$\,ms), an appropriate object for probing  scattering effects given its large $\DM = 297 \, \DMunits$ and its inclusion in  NANOGrav's timing program. 
Large scattering is consistent with the large distance to this pulsar
($\sim 6\,$kpc) in the Galactic plane
(Galactic coordinates $l = 37.34^{\circ}, b = -1.01^{\circ}$).
While this MSP contributes little to the sensitivity of NANOGrav's  PTA to long-wavelength gravitational waves, observations  at monthly cadence provide the means for a study of DM and scattering along its line of sight \citep{2023ApJ...951L..10A, geigermemo}.  We also note that it is an object of interest for probing 
the neutron star equation of state and, given its likely origin in a triple stellar system,  for understanding millisecond pulsar formation mechanisms \citep{2008Sci...320.1309C, 10.1111/j.1365-2966.2010.18109.x, 2015arXiv151007015K}. 

In Section 2 we briefly summarize the observational data and then discuss pulse-shape modeling in Section 3.  Section 4 presents results of simulations that illustrate 
the origin of systematic errors in estimates for scattering parameters and arrival times.
This is followed by application of our methods to the PSR J1903+0327 in Section 5. 
Section 6 summarizes our results and our major conclusions. 

\section{Observations}
\label{sec:obs}

In this work, we analyze 5 years of pulse shapes from  PSR J1903+0327  collected in NANOGrav's timing program using the Arecibo Observatory in Puerto Rico 
\citep{2021ApJS..252....4A}. These data include all NANOGrav observations of this pulsar at Arecibo that use the Puerto Rican Ultimate Pulsar Processing Instrument (PUPPI) backend in the 12.5-year dataset, with a few exceptions. Of the 60 'L-wide' receiver observations in the frequency range of 1100-1800 MHz, we discard 4, and of the 63 'S-wide' receiver observations in the frequency range of 1700-2404 MHz, we discard 5. This is due to either dramatic radio frequency interference or a very low signal-to-noise ratio.
We defer analysis of the NANOGrav 15-year dataset for this pulsar until ongoing work to further address time of arrival (\toa) corrections for this and other pulsars is completed. 
Details of the timing program and data products are described in \citet{2023ApJ...951L...9A}.  Here we summarize pertinent details. 
Average profiles with 2048 samples across pulse phase  were obtained as
$\sim 30$\,minute averages ($\sim 800$k\, pulses). 
 The total intensity (Stokes $I$) profiles were computed after coherent dedispersion and gain calibration of the two receiver polarization channels. 

For calculating pulse arrival times, the NANOGrav 12.5-year data set narrowband analysis used fixed templates for each receiver band, with no compensation in the template shape for profile evolution within the band.   A parallel `wideband' analysis yields arrival times through cross-correlation of a two-dimensional (2D) template with each 2D data profile \citep[][]{Pennucci_2019}.   

Here we analyze pulse shapes vs. frequency using a custom analysis that aims to
(a) generate the intrinsic pulse shape, i.e. that emitted by the pulsar before interstellar propagation, in each of several frequency subbands;
(b) identify the pulse broadening functions that best represent scattered pulse shapes;
(c) analyze the frequency dependence of the characteristic pulse broadening time scale, $\taud$, in terms of a power-law scaling in frequency
$\taud\propto \nu^{-X_{\tau}}$ with scaling index $X_{\tau}$;
and
(d) characterize and interpret the epoch dependence of $\taud$ and the scaling index
$X_{\tau}$. 

\section{Profile and scattering modeling}
\label{sec:models}

Observed pulse profiles, $I(t)$,  are the  convolution of the pulse shape intrinsic to the pulsar, $U(t)$, and the interstellar pulse broadening function (PBF),
$\PBF$,
\be
I(t) = a U(t-t_0) * \PBF(t, \tau),
\label{eq:Idata}
\ee
where the asterisk denotes convolution,  $a$ is a scale factor, $t_0$ is the arrival time in the absence of scattering, and $\tau$ is  the characteristic scattering time that we define, as is typical, to be  the $1/e$ width of the PBF.   Apart from the time $t$, all quantities in  Eq.\,\ref{eq:Idata} are frequency dependent.  In particular, the arrival time $t_0$ includes dispersion delays
\be
\tDM  \simeq 4.15\,{\rm ms}\ \nu^{-2} \DM
\ee
for $\nu$ in GHz and \DM\ in standard units ($\DMunits$). For now, we focus on modeling of $I(t)$ at a single frequency and ignore dispersion delays, which will be treated in a later section. 

The PBF shape and the frequency dependence of  $\taud$ depend on the spectrum of electron density variations and on the distribution of scattering electrons along the \los\ \citep[e.g.,][]{1975RSPSA.342..131W,1975ApJ...201..532L,cr98,lr99}
and transverse to the \los\ \citep[][]{2001ApJ...549..997C}.   

The simplest geometry is 
a thin phase screen (thickness $\ll$ source distance) that scatters incident radiation from a point source into a blurred image. Pulse arrival times are directly related to angles of arrival (AoA) $\theta$ as $t \simeq (d \theta^2/2c) (1-s) / s$, where $d$ is the source distance and $s$ is the fractional distance of the screen along the \los.    Some lines of sight appear to conform to thin screen geometry while others require either multiple screens or a thick medium 
\citep[e.g.,][]{2004ApJ...605..759B,2007A&AT...26..597K,2017MNRAS.470.2659G,2024MNRAS.527.7568O}.
For a single screen, the PBF is a simple transformation of the image $I(\theta)$ to time.  

A commonly used PBF  in pulse-shape modeling is the one-sided exponential, ${\PBF}_{\rm e} = \taud^{-1} \exp(-t/\taud) \Theta(t)$ where $\Theta(t)$ is the Heaviside step function \citep[e.g.,][]{2001ApJ...562L.157L, 2004A&A...425..569L,2010ARep...54..433K,2019ApJ...878..130K,2023arXiv230916765S}.  This form results only for a thin screen and if scattering produces a Gaussian shaped image or, equivalently, a Gaussian visibility function $\Gamma(b) \equiv \exp[-\Dphi(b)/2]$ with $\Dphi(b) \propto b^2$, where $\Dphi(b)$ is the phase structure function for phase perturbations produced by the screen at a displacement along the screen of $b$. 

Substantial evidence exists for PBFs that decay more slowly than an exponential which requires that the associated scattered images decay more slowly in angle than a Gaussian 
\citep[e.g.][]{2004ApJ...605..759B,2009MNRAS.395.1391R}.  Corroboration also comes from studies of parabolic arcs in pulsar scintillations, whose extent in secondary spectra are related to the tail of the underlying scattered image \citep[][]{wmsz04,crsc06}.
Use of an exponential PBF for pulse modeling and \toa\ correction  is problematic and generally inappropriate for several reasons that we elaborate on in this paper. 

A variety of observations 
\citep[e.g.,][and references therein]{1995ApJ...443..209A,2019NatAs...3..154L,2021NatAs...5..761O}
favor a broad, power-law wavenumber spectrum
for electron density fluctuations, with 
an outer scale $\louter$ and inner scale $\linner$, 
\be
P_{\delta n_e}(q) = C_n^2 q^{-\beta} e^{{-(q/q_{\mathrm{i}})}^2},
\label{eq:Pne}
\ee
where $q = 2\pi/b$ is the wavenumber in the range $q \ge 2\pi / \louter$, 
$C_n^2$ is the spectral coefficient, and $\beta$ is  the spectral index \citep{lj76, rickett_ism_prop}.
The phase structure function is $\Dphi(b) \propto b^{\beta-2}$ for 
$\linner \ll b \ll \louter$ and $\Dphi(b) \propto b^2$
for $b \ll \linner$; it saturates at $b \gtrsim \louter$. 

Numerous constraints on $\beta$ exist based on scattering and scintillation measurements, some indicating consistency with a Kolmogorov spectrum ($\beta = 11/3$).  However, some measurements of pulse broadening in particular suggest strong departures from  $\beta = 11/3$.  One aim of this paper is to show that many of those apparent departures likely result from use of  an incorrect PBF in pulse shape fitting.    
While there are few constraints on the inner scale, estimated values $\linner \simeq 70-1000$\,km 
\citep[][]{sg90,1995ApJ...438..708M,2004ApJ...605..759B,2009MNRAS.395.1391R} imply that PBFs should depart significantly from an exponential decay.   

Given the wide range spanned by   the inner and outer scales that appears to apply to the ISM (e.g., $\linner \lesssim 10^3$\,km and $\louter \gtrsim 0.01$\,pc), the shallower-than-square-law regime with $\Dphi \propto b^{5/3}$ for $\beta = 11/3$ is relevant 
to many lines of sight at many observing frequencies.   For distant pulsars or pulsars observed at low frequencies, however,  scattering can be dominated by the inner scale and $\Dphi$ will be approximately square law in form.  


PBFs depend on frequency in two ways.   Their characteristic widths $\tau$ vary strongly as approximately the inverse fourth power \citep{sut71,1975ApJ...201..532L}.   At least as  important is the variation of the  {\it shape} of the PBF  for different values of the dimensionless quantity $\zeta \equiv \linner / \ld$, where $\ld$ is the diffraction scale (see Equation \ref{eq:ld_for_zeta}), and thus for different frequencies. For large $\zeta$, the diffraction is determined by the distinct physical, inner scale. For small $\zeta$, however, different scales determine the scattering at different frequencies.
Both $\zeta$ and the wavenumber index  $\beta$ are needed to specify the shapes of   different PBFs. 

\begin{figure}[t!]
\centering
\includegraphics[width = \columnwidth]{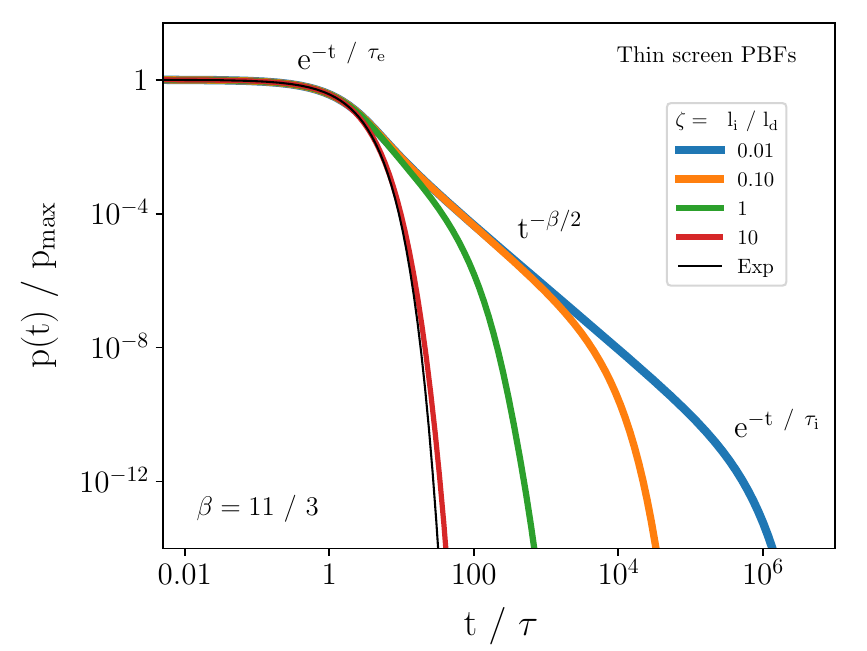}
\caption{PBFs for a power-law wavenumber spectrum with different values of the inner scale along with an exponential PBF.   All PBFs are scaled to unit maximum and have the same $1/e$ time scale, $\tau_e$.   The roll off at larger times $t \sim \tau_i$ is the time at which the power law region of the PBF falls off due to the inner scale.  Though the specific curves are for a Kolmogorov spectrum ($\beta = 11/3$) the power-law
region where $p(t) \propto t^{-\beta/2}$ is labelled generically. 
}
\label{fig:pbfs}
\end{figure}

%
%
The scattered image for $\zeta \ll 1$ rolls off more slowly than a Gaussian function and the PBF correspondingly  extends to
times $t \gg \taud$ \citep{1978RQE...20..581O}.  The PBF is
approximately exponential in form for $t \lesssim \tau_e$ (where $\tau_e$ is the $1/e$ time of the PBF) but has
a power law component $\propto t^{-\beta/2}$ that eventually 
 rolls off exponentially at large
times $t \gtrsim \tau_{\rm i} \simeq (2\pi/\zeta)^2\taud$
(unpublished derivation). This is demonstrated in Fig. \ref{fig:pbfs}. At progressively lower frequencies, $\zeta$ increases and the $t^{-\beta/2}$ regime narrows.  For sufficiently large $\zeta \gg 1$, the PBF tends to a purely exponential form   because the scattering is dominated by the inner scale and the image becomes Gaussian in form.  

The `heavy tailed' PBFs for the power-law regime, which are relevant to nearby pulsars observed at standard frequencies used in PTA observations,   have a strong influence on pulse modeling and arrival time estimation. 

We have generated PBFs for thin screens and thick media over a grid of  $\beta$ and $\zeta$ values.
Fig.\,\ref{fig:pbfs} shows example PBFs for thin screens with different values of
$\zeta$ along with  an exponential PBF.  
   For thin screens it is straight forward to compute phase structure functions, the implied visibility functions and images, and the resulting PBFs using the 
\aoa\ to \toa\ mapping described above.  
  All PBFs are scaled to unity maximum and have the same $1/e$ time.   The log-log plot shows the initial
exponential dependence for $t \lesssim \tau$ and then the low-level but long  power-law tails $\propto t^{-\beta/2}$ for small values of $\zeta$, and the final exponential rolloff related to the inner scale, $\linner$.
The power-law tail narrows and ultimately vanishes for $\zeta \gg 1$.

Thick media require numerical integration similar to the approaches in  
\citet[][]{1975RSPSA.342..131W,1975ApJ...201..532L,1978RQE...20..581O,ishimaru79} 
and \citet[][]{lr99}; our implementation is described elsewhere (JMC, book now in preparation). PBFs for thicker  media  have a slow rise that is related to the thickness of the scattering medium but large-time regions of the PBF are similar to those of thin screens.    In this paper the only thick-media PBFs we make use of are for scattering occurring along the entire \los.

A convenient  scaling of 
the $1/e$ scattering time with frequency is the power law form,
\begin{equation}
\tau (\nu) = \tau_0 \left (\nu/\nu_0 \right ) ^ {-X_{\tau}},
\end{equation}
where $\tau_0$ is the value  at a fiducial frequency, $\nu_0$. 
For a medium with a power law spectrum as in Eq.\,\ref{eq:Pne} with $\zeta \ll 1$  and $2<\beta<4$, the scaling index in the inertial subrange regime (between the inner and outer scales, where there is a turbulent cascade in Kolmogorov theory) is  $X_{\tau} = 2\beta/(\beta-2)$, which gives $X_\tau = 22/5$ for $\beta=11/3$. For $\zeta \gg 1$ or for a uniscale medium with approximately 
a square-law structure function, the index has an  asymptotic value
$X_{\tau} = 4$ \citep[JMC et al., in preparation]{2009MNRAS.395.1391R, 1978RQE...20..581O}.   

We now relate the scattering time $\taud$ (defined as the $1/e$ scale of the PBF\footnote{For an exponential PBF, $\taud$ is the only parameter and it is uniquely defined.   For thick media, a useful (but not unique) definition is to measure the $1/e$ width from the maximum of the PBF.}) to other relevant quantities
in the context of a thin scattering screen. 
The diffraction scale is the characteristic size of an intensity feature in the diffraction pattern measured in the observer's plane transverse to the \los,
\be
\ld  &=& \frac{1}{2\pi \nu} \left(\frac{c\deff}{\taud} \right)^{1/2}
\simeq 	\frac{1.43\times 10^4\,\rm km}{\nu} 
		\left(\frac{\deff} {{\taud}} \right)^{1/2} ,
\label{eq:ld_for_zeta}
\ee
for  $\nu$  in GHz, $\deff$ in kpc, and ${\taud}$  in microseconds.
The effective distance is $\deff = s(1-s)d$ for a thin screen at distance $sd$ from the pulsar. 
The expression for $\ld$ is 
obtained by using its definition  from the  phase structure function, $\Dphi(\ld) \equiv 1$,  and relating it to the  characteristic scattering angle
$\thetad = 1 / k\ld$ ($k=2\pi/\lambda$, which determines the scattering time
$\tau \propto \deff\thetad^2/c$. 
Using $\linner = 10^3 \, {\rm km}\, {\linner}_{3}$, we also obtain
\be
\zeta = \frac{\linner}{\ld} 
\simeq 
0.070 \times (\nu{\linner}_3)
\left ( \frac{{\taud}} {\deff} \right)^{1/2}
\propto \nu^{-2/(\beta-2)} .
\label{eq:zeta}
\ee
The proportionality in Eq.\,\ref{eq:zeta} applies when
the diffraction scale satisifes,
$\ld \gg \linner$.    For the opposite case,  $\zeta \propto \nu^{-1}$.

 \toa\ estimation and determinations of $\tau(\nu)$ require model fitting to measured pulse shapes.   In parallel with Eq.\ref{eq:Idata}, we write the pulse model in a form
emphasizing  that all quantities other than the independent time variable $t$
are unknown and must be measured or estimated, 
\be
\Imod(t) = \ahat \,\Umod(t-\toahat) * \PBFmod(t, \tauhat).
\label{eq:Imod}
\ee
If the intrinsic pulse shape and actual PBF are known, i.e. 
$\Umod = U$ and $\PBFmod = \PBF$,  the three model parameters can be retrieved perfectly (in the absence of additive noise), viz.
$\ahat = a, \toahat = t_0$, and $\tauhat = \tau$.  Conversely,  if one or both of
$U$ and $\PBF$ are imperfectly modeled, all three parameters will be systematically biased.   The consequences for multifrequency observations are that 
the scaling index $X_\tau$ for the scattering time, the \DM, and the \TOA\ (and its correction through removal of dispersion and scattering delays)  will all  be misestimated.   For example, values for $X_\tau$ can show strong, spurious departures  from the $X_\tau = 22/5$ value expected for $\beta = 11/3$ or, in the heavy scattering regime, from
$X_{\tau} = 4$.

\section{Implications of Mismatched Intrinsic and Pulse Broadening Function Shapes}\label{sec:implication_mismatch}

We demonstrate that incorrect assumption of the intrinsic or pulse broadening function shapes results in incorrect determination of $\tau$ and $X_{\tau}$. This is not unexpected, but the bias in estimates for  $\tau(\nu)$ assuming an incorrect intrinsic or pulse broadening function shape is dramatic. 

\begin{figure}[h!]
\centering
\includegraphics[width = \columnwidth]{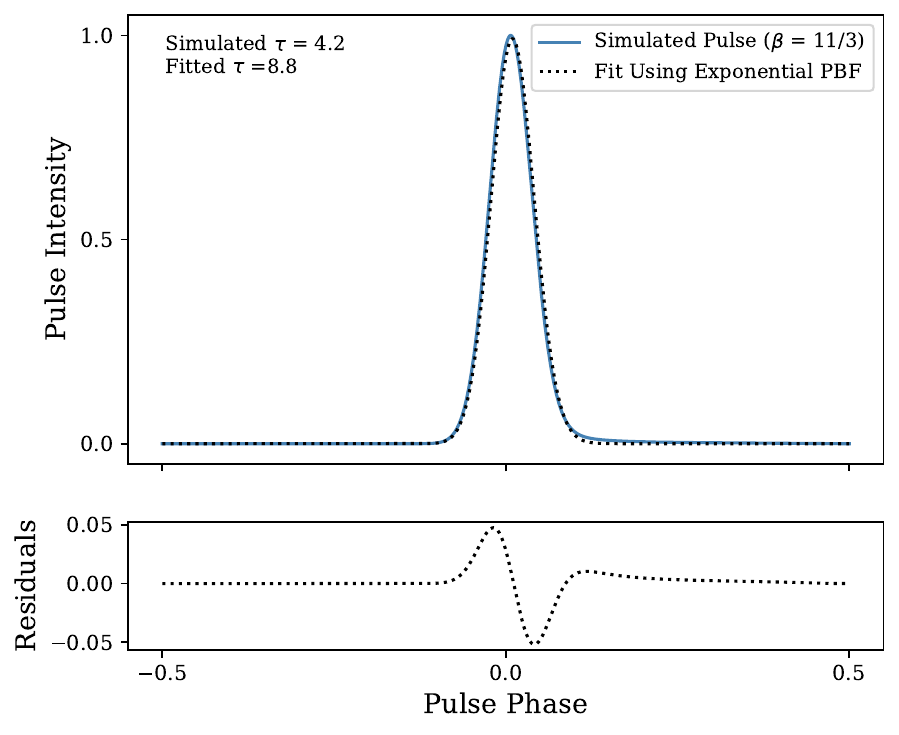}
\caption{Demonstration of incorrect estimation of the pulse broadening time. The  simulated scattered pulse shape is a Gaussian function with
width (FWHM) $W=0.03$ (in pulse phase or cycles  units between 0 and 1) convolved with an extended medium PBF for $\beta = 11/3$, $\zeta \ll 1$, and $\tau = 4.2$\,millicycles.  The fitting function is the same
Gaussian convolved with an exponential PBF.   The best fit $\widehat\tau = 8.8$\,millicycles, more than twice the true value.  The small ratio
$\tau/W \sim 0.13$ yields a barely discernible change in pulse shape by eye but a clear signal in the residuals, which are similar to the derivative of the pulse shape but are asymmetric in time. 
}
\label{fig:singleprofpbfmismatch}
\end{figure}

In Fig. \ref{fig:singleprofpbfmismatch}, we show a simulated
 Gaussian pulse shape scattered by a heavy-tailed PBF for 
 an extended Kolmogorov medium with $\beta = 11/3$ and  $\zeta \ll 1$ (c.f. Fig.\,\ref{fig:pbfs}).  We  model this profile with the same intrinsic Gaussian  pulse but convolved with an  exponential PBF instead of the Kolmogorov PBF.  Estimating
 $\tau$ and the pulse alignment using this incorrect model 
 results in a factor of 2 overestimation of $\tau$ for the particular
 Gaussian width $W = 0.03$\,cycles. Overestimation occurs because the exponential PBF must be broader (at $1/e$) than the true PBF in order to account for the longer tail of the true PBF. 
 Larger values of $W$ yield larger discrepancies between true and fitted $\tau$ values.   The residuals in the bottom panel 
 resemble the derivative of the intrinsic Gaussian shape, indicative of   an overall profile shift. This shift, like the $\tau$ overestimation, also compensates for the shorter exponential tail. 

\begin{figure}[h]
\centering
\includegraphics[width = \columnwidth]{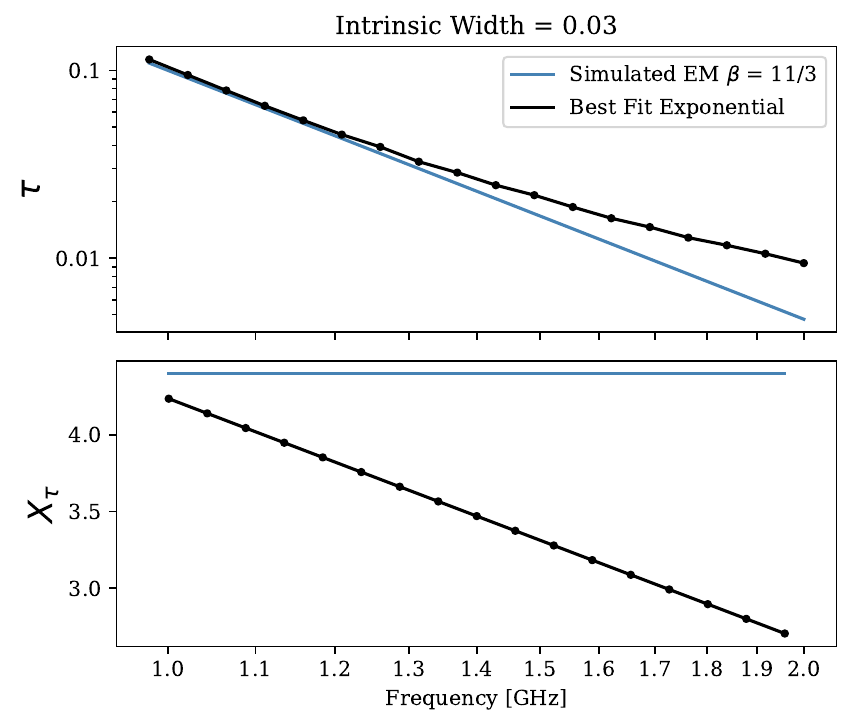}
\includegraphics[width = \columnwidth]{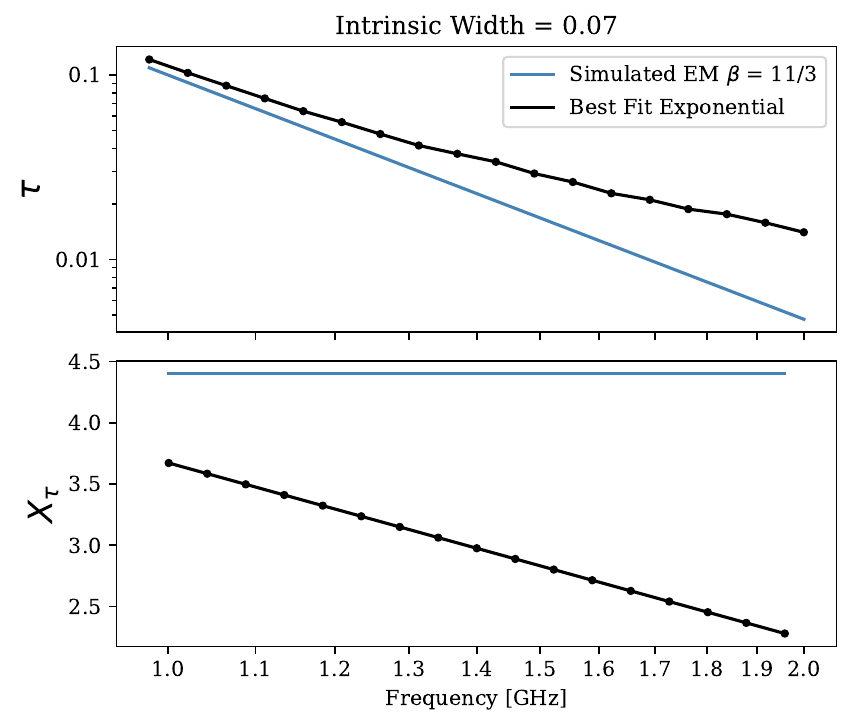}
\caption{Demonstration of estimation biases for $\tau$ and $X_\tau$ vs. frequency
using an incorrect exponential PBF and
two different intrinsic pulse widths. 
Each case considers a simulated profile with a single Gaussian intrinsic shape with widths $W = 0.03$\,cycles (top pair of panels) and $W=0.07$\,cycles (bottom pair).   Scattering is with  an extended medium (EM) PBF with $\beta = 11/3$ and $\zeta \ll 1$.  The true scaling index
for the simulated profiles is $X_{\tau} = 4.4$  and $\tau_0 = 0.1$\,cycles at a 1 GHz reference frequency.  To demonstrate the importance of correctly matching the PBF shape when performing scattering analyses, we fit scattered pulses assuming the correct intrinsic shape and  width but an incorrect exponential PBF. The intrinsic shapes in this simulation do not evolve with frequency. 
To obtain $X_{\tau}$, we fitted a polynomial to each $\tau$ vs. frequency curve and calculated its derivative.}
\label{fig:manyprofpbfmismatch}
\end{figure}

%
Extending these single profile results to the multi-frequency case, the approximation of an extended, Kolmogorov medium with an exponential PBF generally yields values for the scaling index  $X_{\tau}$ that are significantly smaller than the true $X_{\tau}$, i.e. a shallower dependence of $\tau(\nu)$ on frequency, as shown in Fig. \ref{fig:manyprofpbfmismatch}. In these simulations, the intrinsic pulse shape is constant in frequency.  Our results may account for some of the findings in the  literature that report
  shallow $X_{\tau}$ and claim departures from Kolmogorov media, given that most scattering analyses assume light-tailed  exponential PBFs.
Alternative explanations may also apply, such as anisotropic scattering that produces longer-tailed PBFs \citep[e.g.,][]{2009MNRAS.395.1391R,2017MNRAS.470.2659G} or scattering regions that are limited in extent transverse to the line of sight \citep[][]{2001ApJ...549..997C}.

\begin{figure}[h]
\centering
\includegraphics[width = \columnwidth]{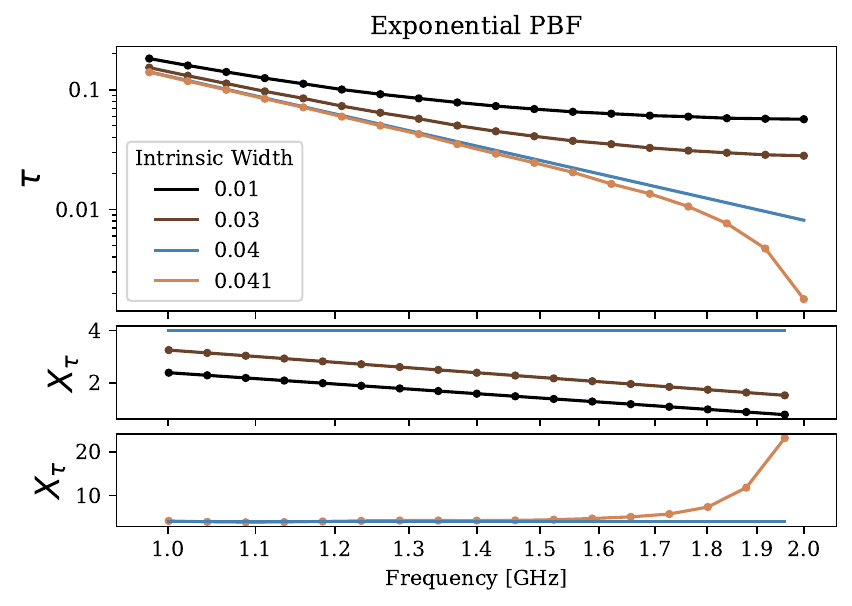}
\caption{Demonstration of biases when using the correct PBF shape but
incorrect intrinsic pulses widths.   Intrinsic Gaussian pulses with 
width $W = 0.04$\,cycles were scattered with an exponential PBF. 
Fitting was done with the correct PBF but with four assumed values
for $W$, three incorrect and one equal to the true value.
The top panel shows $\tau(\nu)$ while $X_\tau$ is shown in the middle and bottom panels (the latter showing the strong bias for $W = 0.041$ that is larger than the true $W=0.04$). 
The blue curves in the three panels correspond to the correct value of $W$ and demonstrate that the assumed frequency scaling with $X_\tau = 4$
is retrieved.}
\label{fig:manyprofwidthmismatch}
\end{figure}

%
We also show in Fig. \ref{fig:manyprofwidthmismatch} that shallow measurements of $X_{\tau}$ can result from underestimation of the intrinsic pulse shape width, hence overestimating the effects of scattering when fitting in order to match the simulated profile width. Similarly, steep measurements of $X_{\tau}$ can result from overestimating the intrinsic pulse shape width, hence underestimating the effects of scattering. However, this seems to be a less common mismatch, since it becomes obvious in high frequency, weakly scattered profiles. We suggest rigorous intrinsic shape and PBF fitting in order to achieve accurate measurement of scattering times.

\section{Application to PSR J1903+0327}

We analyze multiepoch, mulifrequency profiles from PSR J1903+0327 with the following goals:  
(1) develop best-fit models for the intrinsic pulse shape and PBF shape,
$\Umod(t)$ and $\PBFmod(t)$,  both of which are frequency dependent; 
(2) obtain estimates of $\taud(t,\nu)$ vs. frequency and epoch;
and
(3) estimate the scaling index $X_\tau(t)$ vs. epoch.




\subsection{Candidate Pulse Broadening Functions (PBFs)}

In our analysis we used PBFs for power-law media and compared results with those obtained using an exponential PBF. Two sets of power-law PBFs were considered: one with
$\zeta \ll 1$ and $\beta$ = 3.1, 3.5, 11/3, 3.975
and another with fixed $\beta = 11/3$ and $\zeta$ = 0.03 (0.02 for extended medium case), 0.05, 0.5, 1.0, 5.0 for both thin screen and extended media. We did not use a large grid in $\beta$ and $\zeta$ because numerical PBFs for extended media are computationally intensive to generate and because the signal-to-noise ratio of the data does not allow for fine-grained discrimination between models.


%
%


\subsection{Modeling the Intrinsic Pulse Shape}\label{intrinsic modeling}

\begin{figure}[t]
    \centering
    \includegraphics[width=\columnwidth]{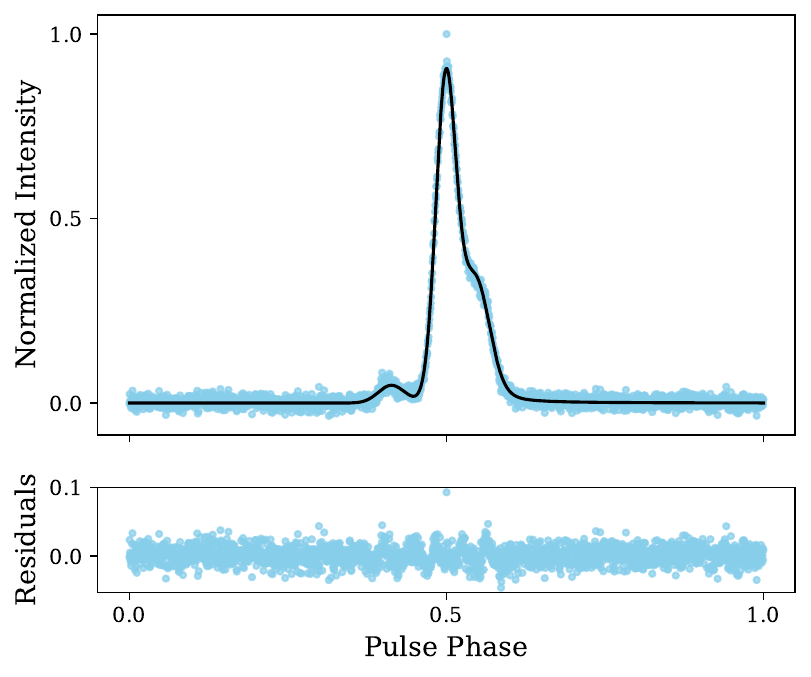}
    \caption{An example fit to the PSR J1903+0327 S-band (2132 MHz) averaged pulse assuming a thin screen PBF with $\beta$ = 11/3 and $\zeta$ = 0.03 and using a best fit model for the intrinsic (emitted) pulse shape.  
    Top panel: measured profile (blue) and fitted model (black) that yielded
    $\tauhat = 11\pm 1\, \mu$s with $\chi^2 = 1.19$. 
    Bottom panel: residuals from the fit. 
    The fitted triple Gaussian intrinsic shape parameters are $\mathrm{A}_1 = 0.053\pm0.002$, $\mathrm{A}_3 = 0.33\pm0.004$, $\phi_1 = 0.41\pm0.001$, $\phi_2 = 0.49\pm0.0003$, $\phi_3 = 0.54\pm0.0004$, $\mathrm{W}_1 = 0.020\pm0.0009$, $\mathrm{W}_2 = 0.015\pm0.0001$, and $\mathrm{W}_3 = 0.20\pm0.0005$, where subscripts 1, 2, and 3 indicate the Gaussian components 1st, 2nd, and 3rd in phase, A indicates amplitude as a fraction of the 2nd component amplitude, $\phi$ indicates phase, and W indicates width as a fraction of pulse phase.}
    \label{fig:sbangavg}
\end{figure}

Incorrect modeling of either the intrinsic or PBF shape can dramatically affect measurements of $\tau$, as demonstrated in
Section \ref{sec:implication_mismatch}. We  model the  intrinsic shape using three components that  are  particularly visible at S-band (1800-2400 MHz),  where scattering is less dramatic compared to lower frequencies
(Fig.\,\ref{fig:sbangavg}). 
We use a triple-Gaussian model,
\be
\Umod(t) = \sum_{j=1}^{3} A_j g(t-\phi_j P, W_j P)
\label{eq:intrinsic_model}
\ee
where $g(x,y) = \exp[-4\ln 2 (x/y)^2]$ with $\phi_j$ and $W_j$ expressed in
phase units (cycles), the width $W_j$ is the full widths at half maximum (FWHM),  and the  amplitudes are relative to  $A_2 \equiv 1$.

We averaged S-band profiles at all epochs  and
we fitted for the relative phases, widths, and amplitudes of the three Gaussian components. At the same time, we fit for the $1/e$ time scale of a PBF, since scattering is still significant in this band (see Fig. \ref{fig:sbangavg}). We repeat this S-band intrinsic shape fitting process for all candidate PBF shapes, naturally getting a different best fit intrinsic shape for each.

\begin{figure}[t]
    \centering
    \includegraphics[width=\columnwidth]{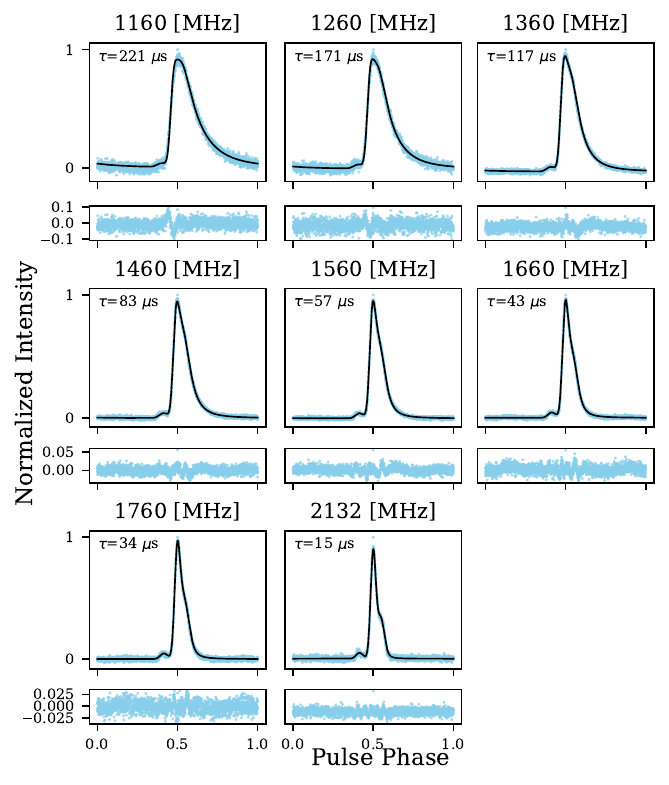}
    \caption{Example fits  (black) to the PSR J1903+0327 frequency subaverages (blue) assuming a thin screen PBF with $\beta$ = 11/3 and $\zeta$ = 1.0. This PBF resulted in the best fit to the frequency subaverages with the average reduced $\chi^2$ of all subaverages being 1.18.} 
    \label{fig:subavg}
\end{figure}

%
We fit for the frequency evolution of intrinsic shape Gaussian parameters using profiles averaged over epoch in 70-MHz wide subbands cross L-band (1150-1800 MHz), as shown in Fig. \ref{fig:subavg}) along with the S-band profile.  We focus on the L-band profiles for this aspect of the fitting because they cover a wider frequency range than the S-band data and scattering is much stronger in the lower part of the band. 
In practice we 
fit for the variation over frequency of the most important four (out of nine) intrinsic shape parameters:  $A_1$,  $A_3$, $W_3$, and $\phi_3$, respectively.
Note that $A_2$ is held fixed and there is an overall scale parameter on all components as part of the fitting process.
These four parameters are the most important because the first component visibly varies in amplitude over frequency, and the third component is most covariant with scattering, warranting more careful modeling. To give a sense for the degree to which the third component and broadening due to scattering blend together, $\tau$ is about 10\% of the width of the third component of the best fit intrinsic shape at 1500 MHz for a thin screen PBF with $\beta$ = 11/3 and
$\zeta$ = 0.03. We model each of these parameters as power laws in frequency,
\begin{equation}\label{Intrinsic Amplitude 1}
R(\nu) = R_{\mathrm{S}} \left(\frac{\nu}{\nu_\mathrm{s}}\right)^{-X_{R}},
\end{equation}
where $R$ = ($A_1$,  $A_3$, $W_3$, or $\phi_3$). In each case, $\mathrm{R}_{\mathrm{S}}$ is the MCMC fitted value of the corresponding intrinsic shape parameter at 
the average S-band frequency  of 2132 MHz. Simultaneously, we allow  $\tau$ to vary freely  at each frequency. We, again, repeat this process for all candidate PBF shapes. The best fit $X_{A_1}$, $X_{A_3}$, $X_{\phi_3}$, and $X_{W_3}$ results for all PBFs are given in Table \ref{table:freqev}. For the remaining parameters not fitted to vary across frequency, we set their values to those best fit at S-band (except for the overall pulse height scale factor which is necessarily free to vary for each pulse fit). From these fits, we get an initial estimate, based on the epoch average profile, of the best fit PBF. This is given by the minimum of the $\chi^2$ values in Table \ref{table:freqev}.

\begin{figure}[h!]
    \centering
    \includegraphics[width=\columnwidth]{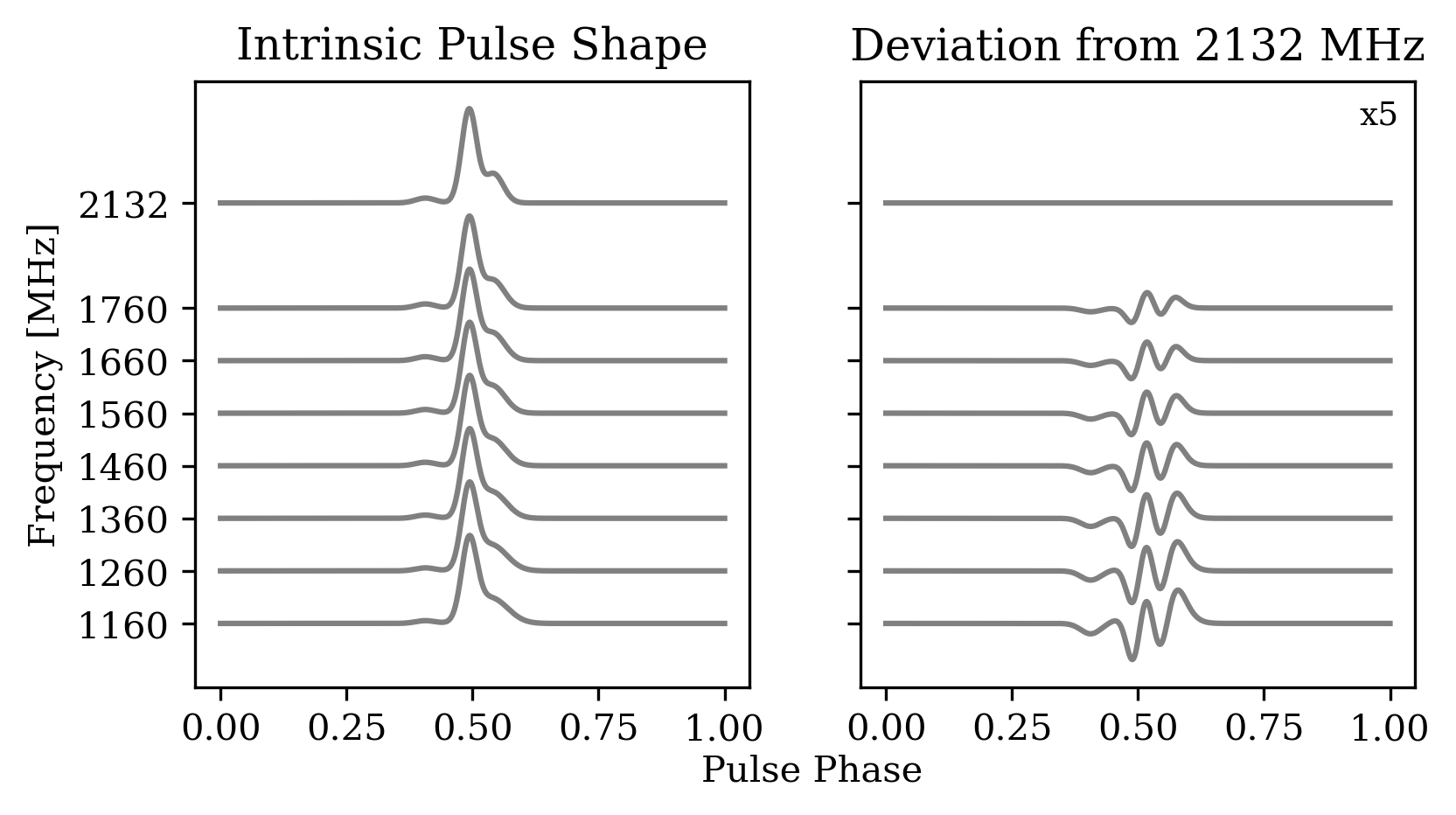}
    \includegraphics[width=\columnwidth]{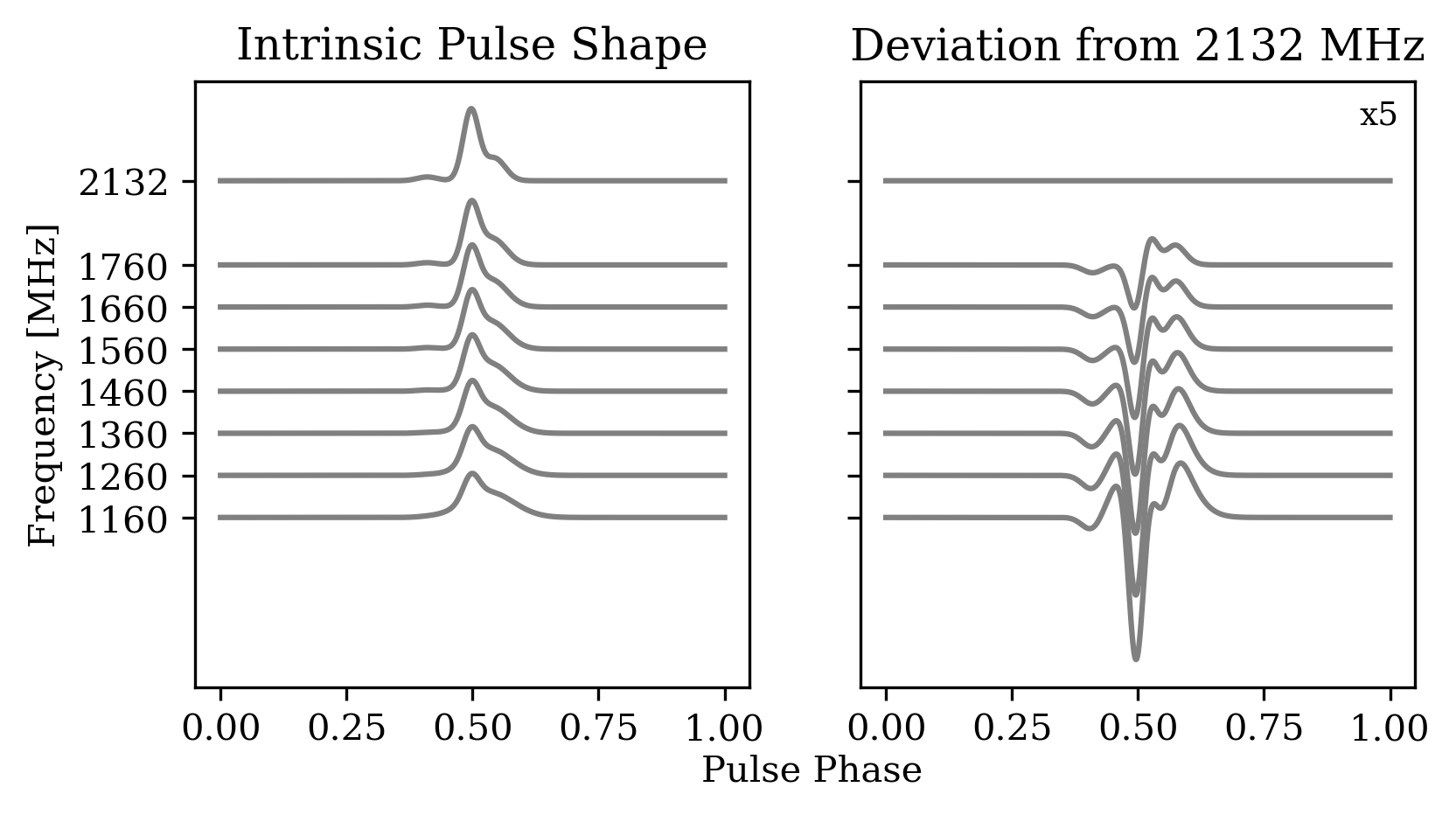}
    \caption{
    Modeled intrinsic pulse shapes (left panels) and their differences (right panels) from the 2132 MHz shape vs. radio frequency.    The top two panels are for an intrinsic 
    shape model derived using the thin screen, $\beta = 11/3$ and $\zeta = 1.0$ PBF as used in Fig.\,\ref{fig:subavg}
   (see caption). For comparison, the bottom two panels show  intrinsic shapes
    derived for an extended medium PBF with $\beta$ = 3.1 and $\zeta \approx$ 0.}
    \label{fig:subavgintrins}
\end{figure}

\begin{figure}[h!]
    \centering
    \includegraphics[width=\columnwidth]{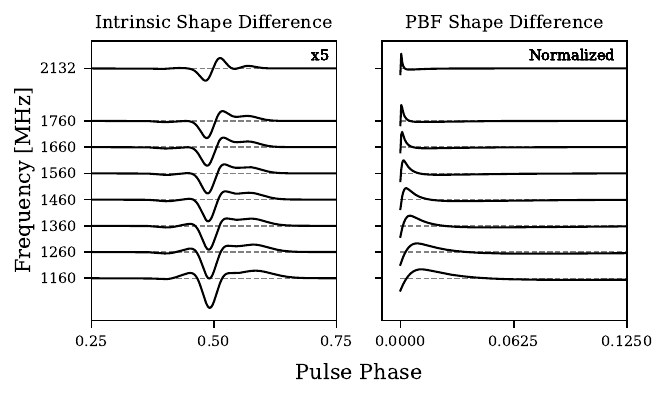}
    \caption{
    A comparison of the fitted intrinsic shapes shown in the top and bottom panels of Fig. \ref{fig:subavgintrins} (see caption). We provide the difference in the corresponding intrinsic and PBF shapes. The intrinsic and PBF shape differences are computed using the unit area normalization of $I(t)$ and $p(t,\tau)$, respectfully, at each frequency. For plotting, the PBF shape differences are normalized by the maximum of the two considered PBFs, and the intrinsic shape differences are given at 5x the scale of the intrinsic shapes if they were normalized to uniform area on these axes.} 
    \label{fig:intrinsdiff}
\end{figure}

We provide an example pulse shape fit across frequency assuming a $\beta = 11/3$, $\zeta$ = 1.0 thin screen PBF in Fig. \ref{fig:subavg}, the corresponding intrinsic shape in Fig. \ref{fig:subavgintrins}, and a comparative intrinsic shape corresponding to assumption of an extended medium $\beta$ = 3.1, $\zeta \ll 1$ PBF in Figs. \ref{fig:subavgintrins} and \ref{fig:intrinsdiff}. The fits shown in Fig. \ref{fig:subavg} have best fit 
 intrinsic S-band shape parameters,   
 $A_1 = 0.053\pm0.002$, 
 $A_3 = 0.33\pm0.004$, 
 $\phi_1 = 0.41\pm0.001$, 
 $\phi_2 = 0.49\pm0.0002$, 
 $\phi_3 = 0.54\pm0.0003$, 
 $W_1 = 0.020\pm0.001$, 
 $W_2 = 0.015\pm0.00009$, and 
 $W_3 = 0.020\pm0.0005$.
 The corresponding frequency scaling indices are
 $X_{A_1}$, $X_{A_3}$, $X_{\phi_3}$, and $X_{W_3}$ are 0.6, $-$0.1, 0.2, $-$0.9, respectively with error of $\pm$ 0.1.

\begin{deluxetable}{c c c c c c c}[b!]\label{table:freqev}
\centering 
\tablecaption{The frequency evolution of the best fit intrinsic shape for PBFs having different $\beta$ and $\zeta$.  All $X_{A_1}$, $X_{A_3}$, $X_{\phi_3}$, and $X_{W_3}$ have error bars of $\pm$ 0.1.}
\tablehead{$\beta$ & $\zeta$ & $X_{A_1}$ & $X_{A_3}$ & $X_{\phi_3}$ & $X_{W_3}$ & $\chi^2$}
\startdata
\multicolumn{2}{c}{Thin Screen} \\
\midrule
3.1 & $\approx$ 0 & 1.5 & $-1.6$ & 0.6 & $-1.3$ & 2.743\\
3.5 & $\approx$ 0 & 0.8 & $-0.8$ & 0.3 & $-1.1$ & 1.463\\
11/3 & $\approx$ 0 & 0.6 & $-0.5$ & 0.3 & $-1.2$ & 1.287\\
11/3 & 0.03 & 0.6 & $-0.5$ & 0.3 & $-1.2$ & 1.287 \\
11/3 & 0.05 & 0.6 & $-0.5$ & 0.3 & $-1.2$ & 1.287\\
11/3 & 0.5 & 0.8 & 0.1 & 0.1 & $-0.7$ & 1.234\\
11/3 & 1.0 & 0.6 & $-0.1$ & 0.2 & $-0.9$ & 1.183\\
11/3 & 5.0 & 0.5 & 0.5 & 0.2 & $-0.7$ & 1.324\\
3.975 & $\approx$ 0 & 0.6 & 0.9 & 0.1 & $-0.2$ & 1.369\\
\multicolumn{2}{c}{Exponential} & 0.6 & 0.8 & 0.1 & $-0.3$ & 1.412\\
\midrule
\multicolumn{2}{c}{Extended Medium} \\
\midrule
3.1 & $\approx$ 0 & 1.6 & $-2.0$ & 0.5 & $-1.4$ & 2.694\\
3.5 & $\approx$ 0 & 1.0 & $-1.1$ & 0.3 & $-1.3$ & 1.671\\
11/3 & $\approx$ 0 & 1.0 & $-0.2$ & 0.1 & $-1.1$ & 1.508\\
11/3 & 0.02 & 1.0 & $-0.2$ & 0.1 & $-1.1$ & 1.502\\
11/3 & 0.05 & 0.6 & $-0.5$ & 0.3 & $-1.2$ & 1.485\\
11/3 & 0.5 & 0.8 & 0.1 & 0.1 & $-0.7$ & 1.213\\
11/3 & 1.0 & 0.6 & $-0.1$ & 0.2 & $-0.9$ & 1.360\\
11/3 & 5.0 & 1.5 & $-0.4$ & $-0.5$ & 3.2 & 1.611\\
3.975 & $\approx$ 0 & 1.3 & 0.1 & $-0.4$ & 1.9 & 1.570\\
\enddata
\end{deluxetable}
%

%
%
%

\subsection{Estimation of $\tau$ and the frequency index $X_{\tau}$ }\label{sec: tau and x tau estimation}

As with the simulations shown in Section \ref{sec:implication_mismatch}, we found that some PBFs and intrinsic shapes, particularly the exponential PBF, yielded variations of $\tau(\nu)$ with frequency that deviated from a simple power law, with upturns or downturns at the lower frequencies.  We thus favor the shapes that yield a consistent power law variation at all frequencies.   This Occam's razor inference also yields a frequency scaling that is largely consistent with the Kolmogorov scaling in the inertial subrange regime, $\tau(\nu) \propto \nu^{-4.4}$.

\begin{figure}[h!]
    \centering
    \includegraphics[width=\columnwidth]{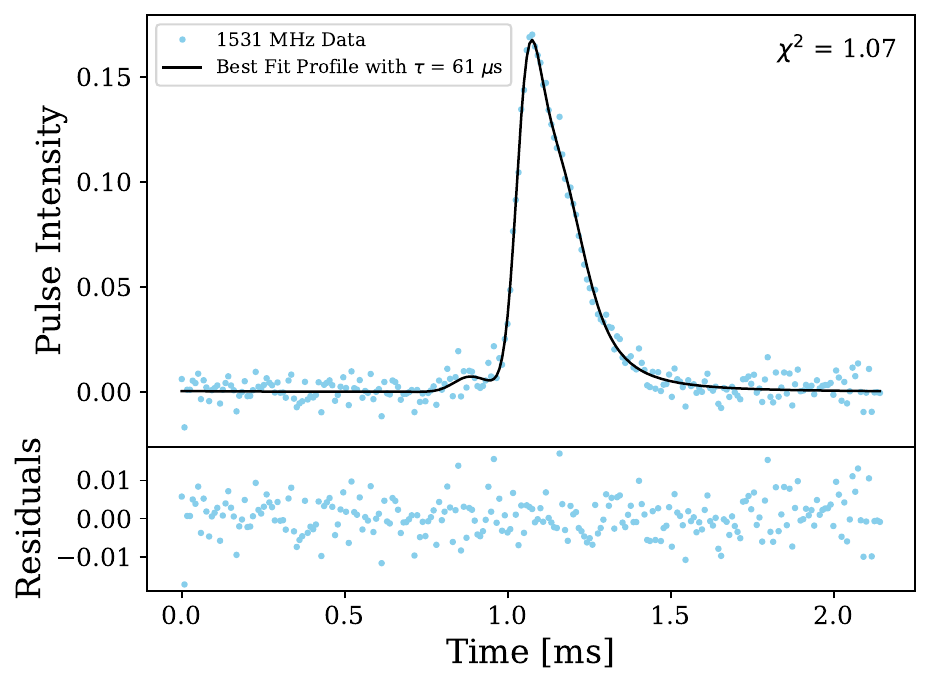}
    \includegraphics[width=\columnwidth]{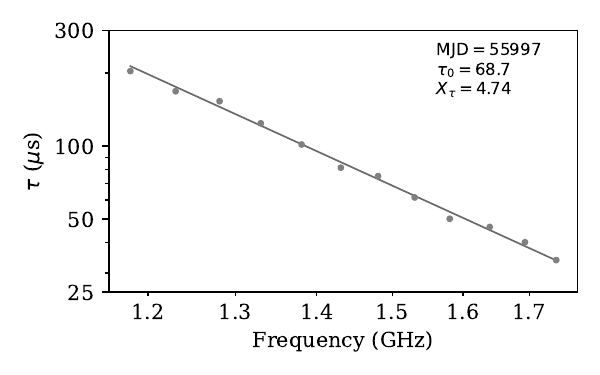}
    \caption{
    Top panel:
     an example fit for $\tau$ at 1531 MHz on MJD 55997 assuming a thin screen PBF with $\beta$ = 11/3 and $\zeta$ = 1.0. The resulting best fit value is $\tau = 61\pm1.9\mu$s.
    Bottom panel: scattering time vs. frequency using the best fit intrinsic
    shape model along with a thin screen PBF with $\beta$ = 11/3 and $\zeta$ = 1.0. A power-law fit to $\tau$ yields  a 
    frequency index $X_\tau = 4.74\pm0.08$ and $\tau = 68.7\pm0.7\mu$s at 1.5\, GHz. 
    The corresponding value at 1\,GHz is 469$\mu$s. 
      Error bars are included in the bottom panel but are smaller than the data point size.} 
    \label{taucalc}
\end{figure}

Setting a PBF and the corresponding intrinsic shape independent of epoch, we fit for $\tau$ at each frequency and epoch using least-squares model fitting to profiles obtained in 70-MHz subbands. 
We then fit for $X_{\tau}$ using multiple frequency values of $\tau$  at each of the 56 epochs of data. For each value of $\tau$, we include the error due to the finite scintle effect \citep{1986ApJ...310..737C}. An example $\tau$ measurement and an example $X_{\tau}$ measurement are provided in Fig. \ref{taucalc}. This process of fitting to each frequency channel profile of each epoch is repeated for each PBF shape (with the corresponding best fit intrinsic shape for said PBF). This allows us to compare the performance of each PBF based on the average of the $\chi^2$ values of each profile fit using the PBF. This is more robust than the estimate of the best fit PBF described in Section \ref{intrinsic modeling}, which considers only the fit to the epoch-averaged profile. In both cases, the lowest $\chi^2$ value corresponds to a thin screen case with $\beta = 11/3$ and $\zeta = 1.0$. A $\zeta$ of 1.0 corresponds to an $\linner$ of about 1400 km, using Equations \ref{eq:ld_for_zeta} and \ref{eq:zeta} with $d = 6.4 kpc$ \citep{10.1111/j.1365-2966.2010.18109.x}, $\tau = 71.9\ \mu$s at 1.5 GHz, which is the average value of $\tau$ at this frequency measured using this PBF, and assuming $s = 1/2$. The error on this value for the inner scale is determined by the inner scales corresponding to the neighboring values of $\zeta$ considered within our family of thin-screen PBFs. These are $\zeta = 0.5$, corresponding to an $\linner$ of about 700 km and $\zeta = 5$, corresponding to an $\linner$ of about 6200 km. Overall, the thin screen PBFs better match the data in comparison to the extended medium PBFs. No one PBF performs significantly better than the others, but non-negligible $\zeta$ and approximately Kolmogorov $\beta$ are favored. An independent measurement of $\beta$ can be derived by fitting a power law to the broadening tail of a pulse, which follows a \(t^{-\beta/2}\) curve \citep{1978RQE...20..581O}. With the signal-to-noise ratios of the profiles in this dataset, this method provides a weak constraint on $\beta$, but a constraint nonetheless consistent with a Kolmogorov wavenumber spectrum.

\subsection{Epoch Dependence of Scattering Parameters}

\begin{figure*}[!ht]
    \centering
    \includegraphics[width=\columnwidth*11/10]{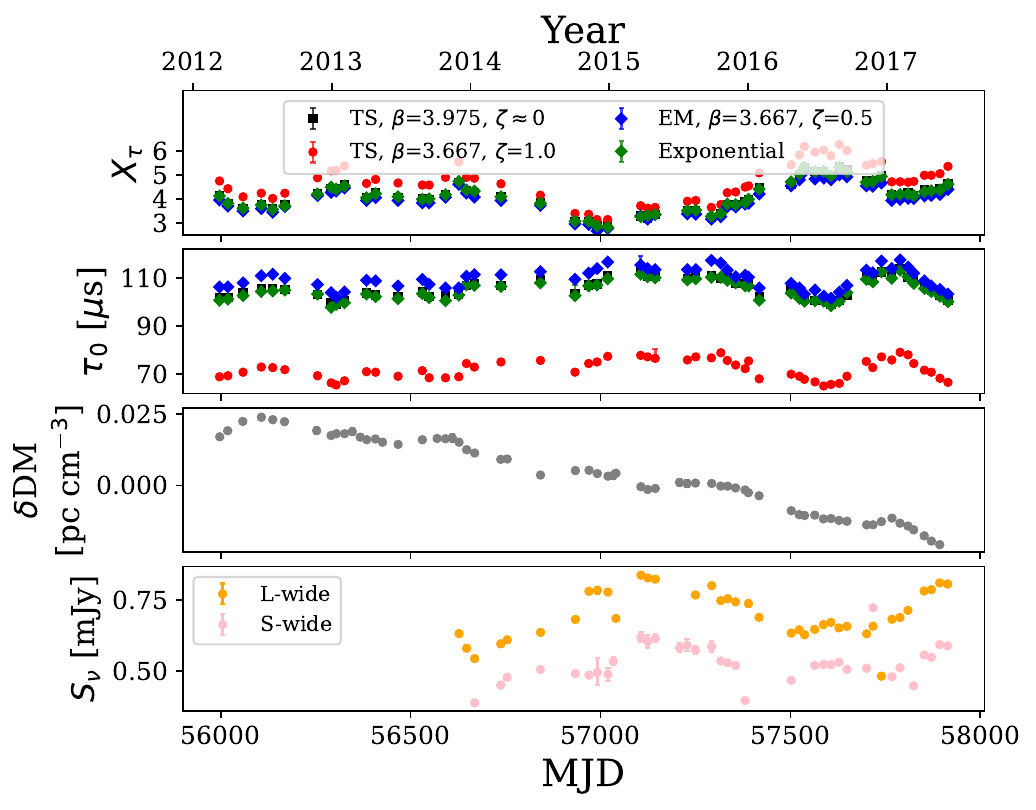}
    \includegraphics[width=\columnwidth]{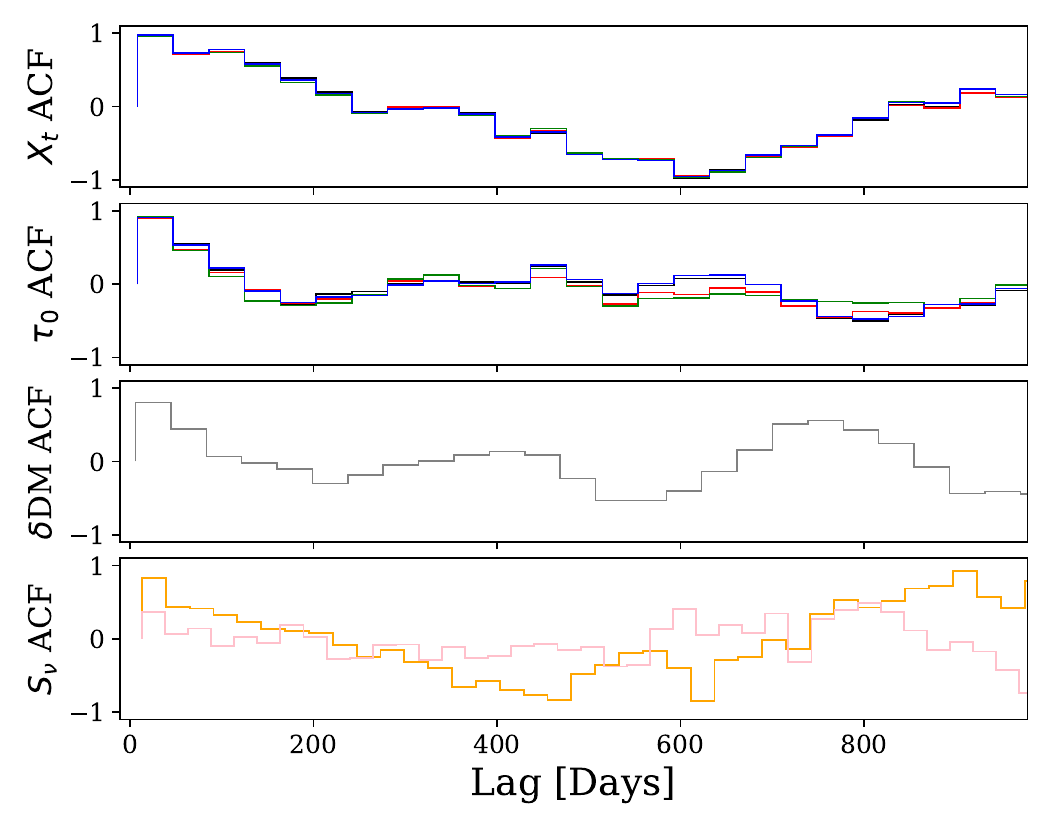}
    \caption{
    Epoch dependence of propagation quantities.  Left: Time series  of the scaling index, scattering time, and $\delta\DM$ using four different PBFs.  Black indicates a thin screen, $\beta$ = 11/3, $\zeta \approx$ 0 PBF, red indicates a thin screen, $\beta$ = 11/3, $\zeta$ = 1.0 PBF, blue indicates an extended medium, $\beta$ = 11/3, $\zeta$ = 0.5 PBF, and green indicates an exponential PBF.  $\tau_0$ is calculated at 1.5 GHz. We provide the variation of DM from a fiducial value of 297.506 pc cm$^{-3}$ over time as observed by NANOGrav for this pulsar for comparison 
    \citep{2021ApJS..252....5A} as well as daily flux density data, $S_{\nu}$, as described in \S 8.1 of \citet[]{2021ApJS..252....4A}.
     Right:  Autocorrelation functions for the time series in the left panels. These data are binned by lag because of the uneven spacing in time between observations. For the $\delta$DM ACF, we first subtract the best fit linear trend from the time series before computing the ACF.} 
    \label{fig:resulttimeseries}
\end{figure*}

Fig. \ref{fig:resulttimeseries}  shows time series  for $\tau_0$ at a fiducial frequency of 1.5 GHz and the scaling index  $X_{\tau}$ based on four different  PBF forms that  either fit the data well or provide a meaningful comparison. For the best fit PBF, a thin screen case with $\beta = 11/3$ and $\zeta = 1.0$, the mean scattering time is $\tau = 503 \pm 19\, \mu$s at 1 GHz, $170 \pm 3\, \mu$s at 
1.25 GHz, 71.9 $\pm$ 0.5$\mu$s at 1.5 GHz, and 35.2 $\pm$ 0.8$\mu$s at 1.75 GHz, and the mean value of $X_{\tau}$ is 4.7 $\pm$ 0.1. We also show the autocorrelation functions for each time series. We find $X_{\tau}$ and $\tau_0$ to be highly variable with long correlation times that are  independent of the assumed PBF. The characteristic time scale for variations in $\tau_0$ and DM is about 100 days, and variations in $X_{\tau}$ have a characteristic timescale of about 250 days as inferred from the autocorrelation functions.  The similarity between the characteristic time scales for $\tau_0$ and DM is expected. From the averaging that occurs over the bundle of rays that reach the observer, a smoothing time scale is imposed on both of these quantities. 
In Fig. \ref{fig:resulttimeseries}, we also provide daily flux density data, $S_{\nu}$, as described in described in \S 8.1 of \citet[]{2021ApJS..252....4A} for comparison (beginning at MJD 56619 due to calibration issues prior to this date), which also has a characteristic timescale of about 250 days. For both L-wide and S-wide, these flux density variations have a modest modulation index, \(m_I \equiv \sigma_{S_\nu} / \langle S_\nu \rangle\), of 0.12. Sizeable uncertainty in these timescales results from the duration of the time series being only a few times the length of these characteristic timescales, particularly for the case of $S_{\nu}$. We note that another heavily scattered pulsar, J1643$-$1224, also shows substantial variation in $\tau$ and $X_{\tau}$ over a 2-year period, not unlike what we see for PSR J1903+0327 \citep{2023arXiv230916765S}. This emphasizes the importance of analyzing and understanding this variation, since it seems to be common for heavily scattered lines of sight.

%
Although variable by about $\pm30$\% (trough to peak), the average  value of the index, $X_{\tau} = 4.7$ with RMS $= 0.8$ for the thin screen, $\beta = 11/3$, $\zeta = 1.0$ best fit PBF is consistent with that expected from a power-law  wavenumber spectrum with the Kolmogorov index $\beta = 11/3$.   A very small inner scale
would yield $X_{\tau,\rm Kol} = 2\beta/(\beta-2) = 4.4$ while an inner scale corresponding to $\zeta > 1$ gives  $X_{\tau, \rm Kol} = 4$.

Variations in $\tau_0$ and $X_\tau$ (and the underlying value of $\beta$) are not expected
to change significantly over the $T = 5$\,yr of data. The transverse offset
of the line of sight over this time is $\lesssim vT \sim 100$\,au, much smaller than the scale sizes of individual structures in the ISM and much smaller than typical  estimates for the outer scale of the wavenumber spectrum.  Lower limits on the outer scale $\gtrsim 100$\,au are implied by DM time series for pulsars that have been monitored for 
20\,yr or longer \citep[e.g,.][]{10.1093/mnras/sty2905,vivekanand2020}.  The existence of interstellar structures (clouds and filaments) larger than pc suggests much larger outer scales, though they could also be line-of-sight dependent.

\section{Modulation of scattering by refraction}

A plausible explanation for  the  variability in $X_{\tau}$ and $\tau_0$ with epoch is  refraction from interstellar structures with $\gtrsim$\,au scales.  Refraction causes slow intensity scintillations, and it also modifies the  shape of the PBF.   For a thin screen, the PBF is directly related to the scattered image, which is distorted and shifted from what it would be in the absence of refraction.   The characteristic time scale is expected to be similar to that of refractive intensity variations. 
This aligns with the observed timescales for $X_{\tau}$ and $S_{\nu}$, with $S_{\nu}$ corresponding to refractive flux density variations because pulsars have been shown to have intrinsically stable flux \citep{1990ApJ...352..207S}.

To derive a relation between the refractive time scale and other observables, we consider the simple case of a thin screen  at distance $sd$ from a pulsar at distance $d$. The variability time scales for refractive and diffractive scintillations are $\dtr = \lr/v$ and $\dtd = \ld/v$, respectively, 
where $v$ is a characteristic transverse velocity that we represent with a fiducial value of 100\, km\,s$^{-1}$.   This velocity is a combination of 
the velocity of the pulsar, the Earth, and the intervening ISM
\citep[e.g.,][]{cr98}. 
The corresponding length scales $\lr$ and $\ld$ are related as
 $\lr \ld = \rF^2$ where the Fresnel scale is 
 $\rF = \sqrt{s(1-s)c d/2\pi\nu}$.  This gives
 $\dtr = v^{-1} \sqrt{s(1-s) cd\tau}$, which we evaluate using
 $\tau = 100\,\mu s$ at 1.5 GHz and the pulsar's estimated distance $d \simeq 6$\,kpc, 
 \be
 \dtr(1.5\,{\rm GHz}) \simeq 136\,{\rm d} \, 
    \left(\frac{100\, \rm km\, s^{-1}}{v}  \right)
    \left[\frac{s(1-s)}{1/4} \right]^{1/2},
 \label{eq:tr}
 \ee
 where the quantity in square brackets applies to a screen midway from the pulsar ($s=1/2$); the refraction time scale scales with frequency as $\dtr \propto \nu^{-\beta/(\beta-2)} = \nu^{11/5}$ for $\beta = 11/3$.  The nominal coefficient is consistent with the apparent time scale for variations in $X_\tau$ and $\tau$ even with nominal (but reasonable) values for $s$ and $v$. 

%
%

Rapid scintillations occur on  a  time scale $\dtd \sim \ld / v \lesssim 100$\,s. as the diffraction pattern moves across the line of sight. However, refractive effects occur   on the much longer  time scale of Eq.\,\ref{eq:tr}, including refractive intensity scintillations \citep[e.g.][]{rickett_ism_prop} along with variations in  \aoa\,, overall image shape, and PBF width and shape, including  variations in the characteristic $1/e$ scattering time $\tau$.   This  stochasticity in $\tau$  is accompanied by variations in \DM. The time scale for \DM\ variations is longer than for the other quantities because $\delta\DM$ is proportional to the screen phase while
the other quantities are related to the first and second derivatives of the screen phase, which enhance high frequency components and de-emphasize low frequency components of the phase.

We simulated refractive effects using a thin phase screen  generated in accordance with a  power-law wavenumber spectrum.  The screen  includes only large refraction scales  by using an exponential rolloff factor  on the power-law spectrum, $\exp(-q / q_r)$ where $q_r = 2\pi / \lr$.   We evaluate the phase and its derivatives  at the emergent point from the screen to calculate the refractive `gain' that modulates the pulsar intensity and the offset, shear, and convergence that shift and distort the image comprising a  diffractive ray bundle.  In the absence of refraction, the scattering disk would be circularly symmetric (in the mean) under the assumption that diffracting irregularities in the screen are isotropic.  Diffraction from anisotropic density fluctuations could of course have been assumed, but the additional complexity was not warranted in our analysis.   The screen generation method is similar to that used in \citet[][]{CSS16} and details of the specific computation reported here will described elsewhere (JMC et al., in preparation).

\begin{figure*}[!ht]
    \centering
    \includegraphics[width=\columnwidth]{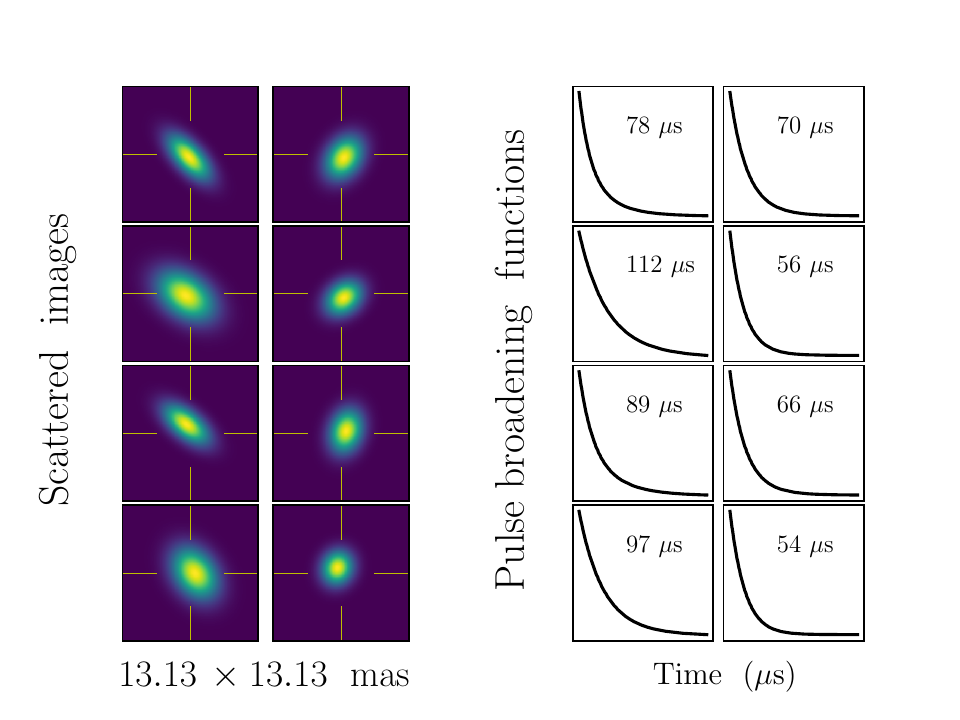}
    \includegraphics[width=\columnwidth]{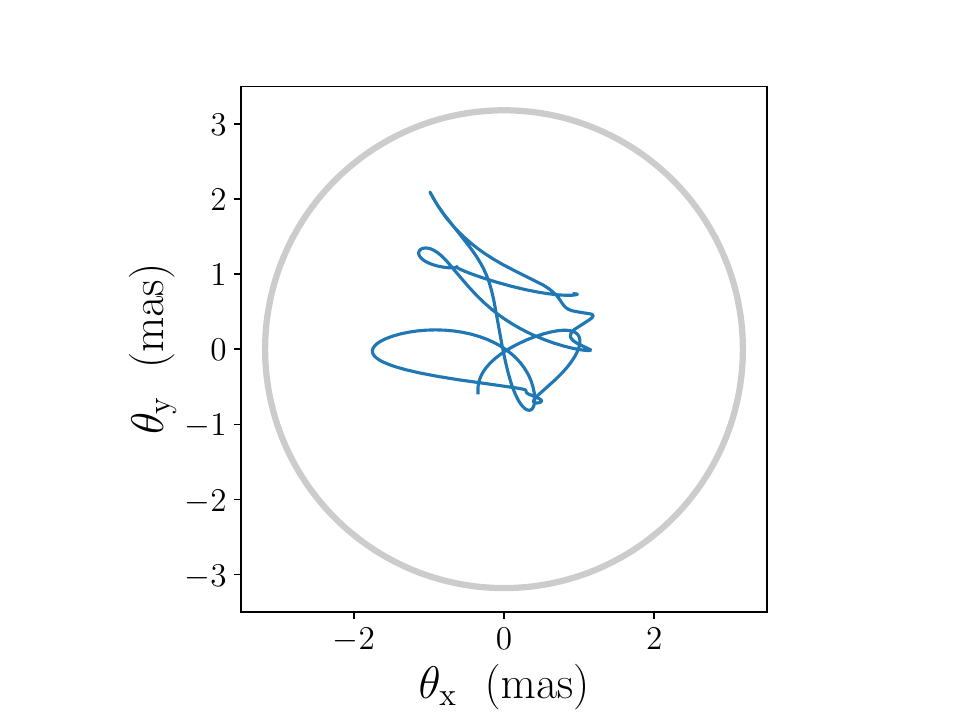}
    \caption{
    Simulated refraction of scattered images from a phase screen with parameter values applicable to the line of sight for PSR J1903+0327. 
    Left: Image and PBFs at different positions along a scattering screen as would be seen at different epochs.[Need to explain the angular scale and time scales; variations in images and PBFs.]
    Right: Wander of the image centroid over a time range corresponding 
    to  about 8\,yr.
    } 
    \label{fig:ref1}
\end{figure*}

Fig.\ref{fig:ref1} shows results for line-of-sight parameters similar to those for PSR J1903+0327
with $\DM = 300\,\DMunits$ and for a 2.1\,GHz radio frequency. 
Scattered images (left panel) are shown  at eight positions along the observation plane with accompanying PBFs (center panel);  refractive wandering of the image centroid is displayed in the right panel.  Without refraction, the images would be nearly  circular and, along with the PBFs, would vary negligibly across the observation plane with much smaller image centroid wandering.
Instead, the images have significant eccentricities and the PBFs vary in shape with $1/e$ $\tau$ values that vary by amounts similar to those seen for PSR J1903+0327 (Fig.\,\ref{fig:resulttimeseries}).   Image wandering in the right-hand panel is smaller than the size of the average scattering disk (black circle) as is consistent for Kolmogorov type media with small inner scales \citep[e.g.][]{rickett_ism_prop}.

\begin{figure}[!ht]
    \centering
    \includegraphics[width=\columnwidth]{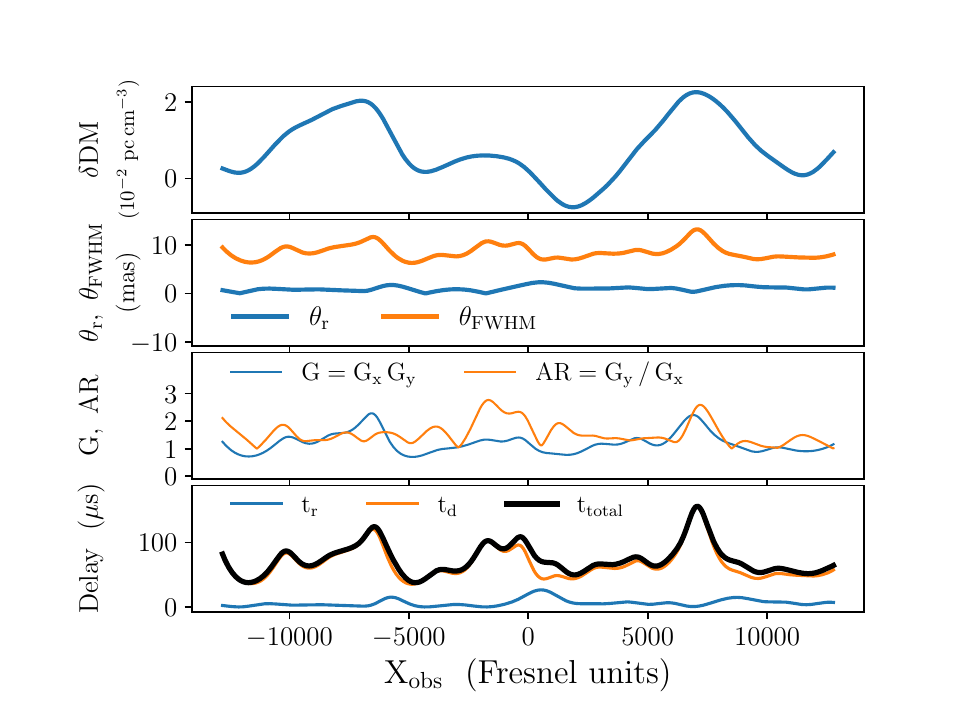}
    \caption{
    Time series of refracted quantities.  From top to bottom:
    $\delta\DM$, the change in dispersion measure;
    the change in image centroid angle and width (FWHM);
    the refraction gain $G$ and the image axial ratio AR;
    and timing delays, including the extra delay $t_{\rm r}$ from the image offset, the diffraction time scale $t_{\rm d}$, and their sum (black curve).  
    } 
    \label{fig:ref2}
\end{figure}

Fig.\,\ref{fig:ref2} shows simulated time series for $\delta\DM$ and scattering angles and time scales along the observation plane measured in units of the Fresnel scale.  For PSR J1903+0327, the Fresnel
scale at 2.1\,GHz for a midway screen is $\rF \simeq 10^6$\,km  corresponding to about 3\,hr for an effective velocity of 100\,km\,s$^{-1}$.  The span of the horizontal axis then represents about 8\,yr, somewhat longer than the 5-yr span for the real data shown in Fig.\,\ref{fig:resulttimeseries}. The top panel shows  variations in $\delta\DM(t)$ that are slower than those seen for the image size (FWHM) and wandering angle (second from top), the intensity gain 
modulation ($G$) and image axial ratio(AR) (third panel), and timing delays shown in the bottom panel. All of the quantities in the bottom three panels are related to derivatives of the screen phase and thus vary faster than $\delta\DM$, which is proportional to the screen phase.   The  time delays in the bottom panel include  the diffractive delay calculated as the mean over the PBF,  $t_{\rm d} = \int dt\, t \,\PBF(t)$ with the PBF $\PBF(t)$ normalized to unit area and the refractive delay 
$t_{\rm r} \simeq  \deff\theta_{\rm r}^2 / 2c$ where $\theta_r$ is the offset of the scattered/refracted image from direct propagation and $\deff = s(1-s)d$ where $s$, as before, is the fractional distance of the screen from the source.  The black curve is the sum of the two delays.  While the goal of the simulation is to simply demonstrate that refraction can cause variations in observable quantities, the chosen screen parameters yield results that are not dissimilar to those seen from PSR J1903+0327.  

\subsection{Pulsar Timing Implications}\label{residual improvement}

We have shown that pulse shape fitting using mismatched shapes for either the intrinsic (unscattered) pulse shape or the PBF yield incorrect estimates for the $1/e$ scattering time $\tau$ and for its scaling index with frequency $X_\tau$.    Such fitting will also yield incorrect \TOA\ estimates that propagate into errors in estimates for \DM\ and thus for  DM-corrected \TOAs.


Scattering induces a \TOA\ delay that is the difference between the observed arrival time of the scattered pulse and the pulse arrival time if there were no scattering effects. 
To calculate this desired \TOA\ shift, we cross-correlate the intrinsic pulse shape with the observed pulse shape. The maximum of this cross-correlation function indicates the ToA shift caused by scattering. 

The effects of scattering on \TOAs\ are simplest if the {\it mean-shift} regime applies
\citep[][]{hs08, cs10}, where the shift in \TOA\ is equal to the mean time of
the PBF, i.e. $\langle t \rangle = \int dt\, t\,\PBF(t)$ where $\PBF$ has unit area. This requires the scattering time  to be much smaller than the intrinsic pulse width ($\tau \ll W$) and the PBF needs to decay sufficiently fast that $t \ll W$ applies for all contributing values of the PBF.   The mean-shift regime can apply if scattering is from a thin screen with a large inner scale ($\zeta \gg 1$) or for a particular line of sight having a narrow range of scale sizes that yield a Gaussian image and exponential PBF.  If this is the case, the TOA shift induced by scattering is simply the mean time of the exponential PBF, $\tau$.  For exponential PBFs with large $\tau \gtrsim W$ this does not apply and the corresponding \TOA\ shift is smaller than $\tau$ by an amount dependent upon the shapes and widths of the intrinsic pulse and PBF. 
More importantly,  {\it the mean-shift regime does not apply to heavy-tailed PBFs} like those from Kolmogorov-like media with small inner scales. 
%

\begin{figure}[h!]
\centering
\includegraphics[width = \columnwidth]{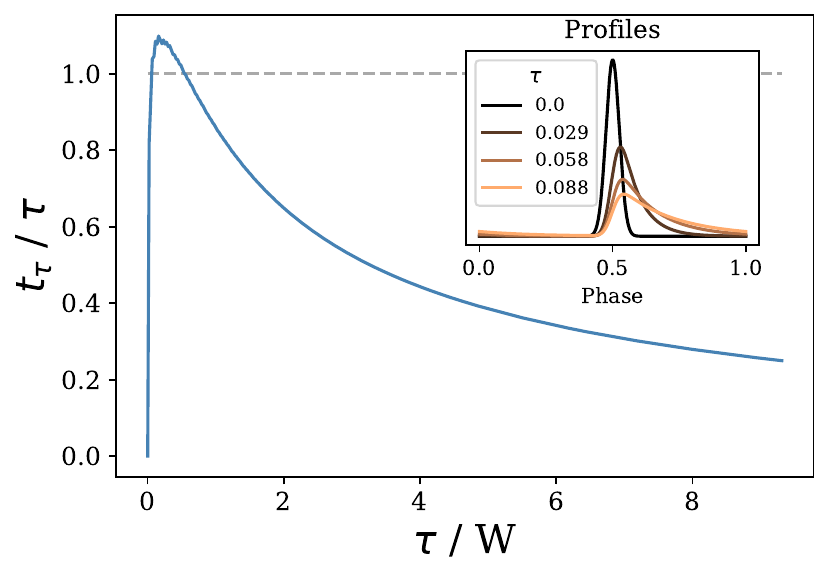}
\includegraphics[width = \columnwidth]{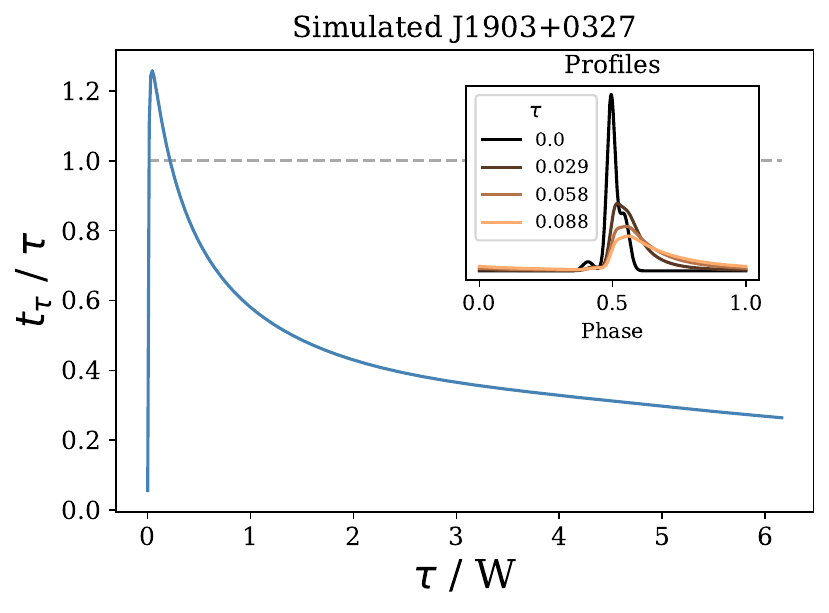}
\caption{The simulated \TOA\ shift, $t_{\tau}$, caused by pulse broadening as a function of scattering time, $\tau$. Top panel: the intrinsic shape assumed is a single Gaussian of width $W$ convolved with an exponential PBF. Bottom panel: the assumed PBF is the PSR J1903+0327 best fit, a thin screen PBF with $\beta$ = 11/3 and $\zeta$ = 1.0. The intrinsic shape assumed is the PSR J1903+0327 S-band best fit for this PBF (see Fig. \ref{fig:subavg}) with FWHM W.  The insets show example scattered profiles for a range of  $\tau$ values.}
\label{fig:timedelay}
\end{figure}

 We now demonstrate the general trend that \TOA\ shifts are $\le \tau$.
 Fig. \ref{fig:timedelay} shows simulated results for the \TOA\ shift induced by scattering as a function of $\tau$; two pulse shapes are shown: (1) a single Gaussian intrinsic pulse shape scattered with an exponential PBF  and (2)  the best fit intrinsic and PBF shapes for PSR J1903+0327. In both cases, for $\tau$ greater than the intrinsic pulse shape width, the induced \TOA\ shift is less than $\tau$. The \TOA\ shift induced by scattering increases at lower frequencies just like $\tau$, but it follows a much shallower trend. In Fig. \ref{fig:dmandscatplotfit} and Table \ref{table:dmmodelparams}, it is shown that for a simulated $X_{\tau}$ of 4.4, the corresponding  \TOA\ shift is modeled well by a power law in frequency with a shallow index of 0.6. This is significantly different from the frequency dependence of $\tau$, and, hence, it is important to consider this true  \TOA\ shift when correcting for scattering. 

\subsubsection{DM Misestimation}

Scattering can cause significant overestimation of DMs if not treated properly.  It is also clear that scattering is epoch dependent and is affected by the non-self similarity of PBF shapes and the changing ratio of scattering time to intrinsic pulse width. 
For $\tau \gtrsim W$, scattering \TOA\ shifts do not follow the simple proportionalities with   $\nu^{-4}$ or $\nu^{-4.4}$, as is demonstrated by the deviation from the line of \TOA\ shifts equal to $\tau$ in Fig. \ref{fig:timedelay}. Scattering \TOA\ shifts follow a shallower function of frequency in comparison to $\tau$, closer to but still steeper than the frequency dependence of dispersion, which is proportional to $\nu^{-2}$. Scattering \TOA\ shifts are therefore absorbed into modern DM measurements and corrections. This results in inaccurate correction of DM and complicates correction of scattering. As scattering changes over time, the amount by which DM is misestimated follows a similar pattern. 

\begin{deluxetable*}{c c c c c}\label{table:dmmodelparams}
\centering 
\tablecaption{The fitted parameters for different scattering and DM corrections corresponding to Fig. \ref{fig:dmandscatplotfit}.}
\tablehead{\colhead{Model} & \colhead{$\delta$DM} & \colhead{$t_{\infty}$} & \colhead{$t_{\tau_0}$} & \colhead{$X_t$}}
\startdata
DM Only & 0.0213 $\pm$ 0.0002 & $-42.3$ $\pm$ 0.5 & - & -\\
DM and Scattering & 0.0 $\pm$ $3\times10^{-5}$ & $-146.0$ $\pm$ 25.2 & 145.1 $\pm$ 25.4 & 0.617 $\pm$ 0.104\\
Scattering Only & - & $-143.7$ $\pm$ 24.4 & 142.8 $\pm$ 24.6 & 0.627 $\pm$ 0.103\\
$X_{t}$ = 4.4, DM Fit & 0.0209 $\pm$ 0.0002 & $-41.4$ $\pm$ 0.5 & $1\times10^{-7}$ $\pm$ 0.086 & -
\enddata
\end{deluxetable*}

We now consider the frequency dependence of \TOAs\ and how different corrections for chromatic terms  influence the net estimate for the achromatic \TOA. 
The measured ToA, $\mathrm{t}_{\nu}$, includes  dispersion and scattering terms,  which we represent as
%
\begin{equation}
t_\nu = \tinfty + \frac{\KDM\mathrm{DM}}{\nu^2} + \ttd\left(\frac{\nu}{\nu_0}\right)^{-X_{t}}, 
\label{eq:dmandscat}
\end{equation}
added to 
 $\tinfty$, the achromatic ToA, figuratively denoted as
 the infinite frequency \TOA\ 
 that would equal the measured \toa\ in the absence of intervening plasmas and additive noise.  
The dispersion constant $\KDM = c r_{\rm e} / 2\pi \simeq 4.15$\,ms, where $c$ is the speed of light, for frequencies in GHz and \DM\ in standard $\DMunits$ units ($r_{\rm e}$ is the classical electron radius). 
The total scattering term, $t_{\tau}$ as in Fig. \ref{fig:timedelay}, equals $\ttd$
at the reference frequency $\nu_0$ and scales with frequency with an index 
$X_t$.
We emphasize  that $\ttd$ is  typically smaller than the scattering time $\taud$ 
and the scaling index $X_t$ is not the same as the index $X_{\taud}$ for the scattering time. Also, while the power-law form is convenient, the scaling index
$X_t$ is itself generally a function of frequency.    

\begin{figure}[h!]
\centering
\includegraphics[width = \columnwidth]{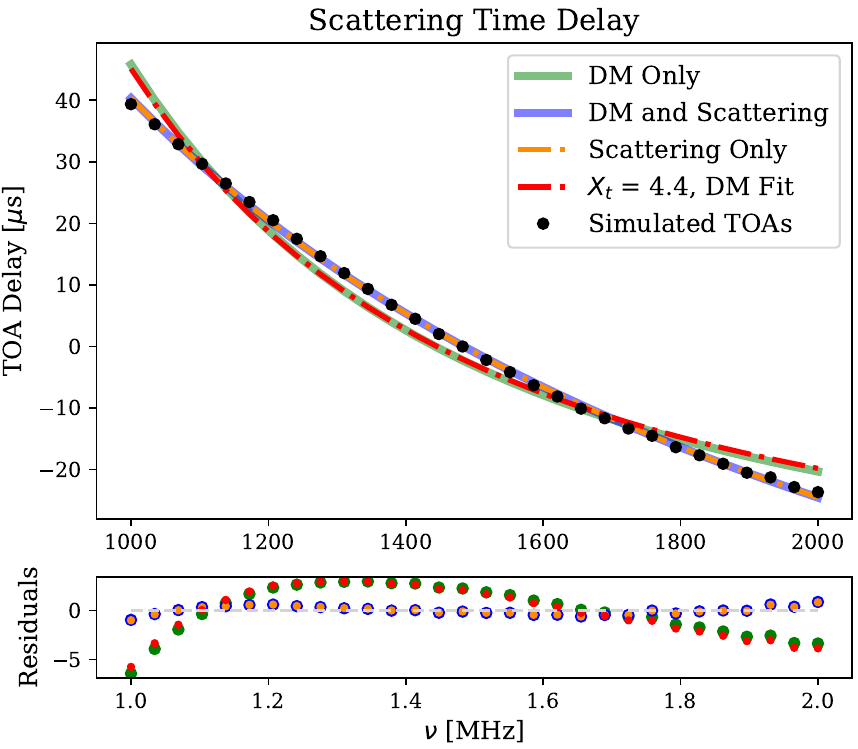}
\caption{
Simulated PSR J1903+0327 scattering \TOA\ shifts, assuming a thin screen, $\beta = 11/3$, $\zeta = 1.0$ PBF, with four different versions of Equation \ref{eq:dmandscat} with $\nu_0$ = 1500 MHz fitted to these simulated \TOA\ shifts. The fit and residuals for each model are given. The ``DM Only'' model ignores scattering, setting
$\ttd$
to zero and fitting for the remaining terms in Eq.\ref{eq:dmandscat}. The ``DM and Scattering'' model includes all terms.  The ``Scattering Only'' model sets $\delta$DM to zero, fitting for the remaining terms, and the "$X_{t}$ = 4.4, DM Fit" fixes $X_{t}$ to 4.4, fitting for the remaining terms. For the resulting fit parameters, see Table \ref{table:dmmodelparams}.}
\label{fig:dmandscatplotfit}
\end{figure}

To demonstrate the interplay between dispersion and scattering, we simulate PSR J1903+0327 scattered profiles assuming a thin screen, $\beta = 11/3$, $\zeta = 1.0$ PBF (the PSR J1903+0327 best fit PBF). We consider a frequency range from 1 GHz to 2 GHz, including intrinsic pulse shape frequency evolution, setting $\tau = 20\%$ of the pulse period at 1 GHz, and setting $X_{\tau} = 4.4$. To best emulate current pulsar timing practices, we calculate the scattering \TOA\ shifts corresponding to these simulated pulses using the 1.5 GHz profile as the template, instead of the intrinsic shape. We then fit variations of the model of $\mathrm{t}_{\nu}$ given in Equation \ref{eq:dmandscat} to the calculated \TOA\ shifts. These models include one with only the DM term, one with both the DM and the scattering terms, one with only the scattering term, and one with both the DM and the scattering terms with $X_t$ set to 4.4. The results are shown in Fig. \ref{fig:dmandscatplotfit} and Table \ref{table:dmmodelparams}. 

Notice that for the ``DM and Scattering" model, a $\delta$DM of zero is favored, matching the ``Scattering Only" model. This is only the case when $X_{t}$ is fitted for. When $X_{t}$ is set, as discussed more below, dispersion only is favored. This is because the simulated scattering \TOA\ shifts are best modeled by a power law with frequency $\propto \nu^{-0.6}$ as in these two models. Similarly, if $X_{t}$ is set to 4.4, a $\mathrm{t}_{\tau_0}$ of zero is favored, corresponding to dispersion only. The frequency dependence of the simulated scattering \TOA\ shifts is neither $\propto \nu^{-4.4}, \nu^{-4},$ nor $\nu^{-2}$. However, its frequency dependence is closest to $\nu^{-2}$, that of dispersion. When fitting only a dispersion term to these scattering \TOA\ shifts (which parallels modern, common practice), the resulting overestimation of DM is 0.0213 $\pm$ $0.0002$ pc/cm$^3$, which is comparable to the magnitude of variations in DM in Fig. \ref{fig:resulttimeseries}.

%


\section{Summary \& Conclusions}

Studies of scattered pulse shapes and  arrival time estimation require careful attention to models for both the emitted (intrinsic) pulse shape $U$ and the pulse broadening function $\PBF$ imposed by multipath propagation.   Using simulations, we showed that strong biases in DMs, scattering times, and \TOAs\ result if either $U$ or the PBF depart from actual shapes.  
In addition,  the frequency dependence of scattering times
$\taud(\nu)$ is also biased, leading to incorrect scaling indices 
$X_\tau = - d\ln \tau/ d\ln\nu$. The often used exponential PBF, 
$\PBF(t) = \tau^{-1} \exp(-t/\tau)\Theta(t)$, is a notable case because its use overestimates $\tau$ when applied to cases where scattering is from a heavy-tailed PBF like those that arise from density fluctuations with a power-law wavenumber spectrum and small inner scale.  The Kolmogorov spectrum is of this type and pulsar measurements indicate that the inner scale in the ISM is quite small, $< 10^3$\,km.
We conjecture that at least some of the reported departures from the Kolmogorov scaling index ($X_\tau = 4.4$ for small amounts of scattering or 
$X_\tau = 4$ for large scattering)   in the literature arise from this issue.

Our detailed analysis of the large-DM PSR J1903+0327  was motivated in part  by it providing an opportunity to study ISM scattering with a well sampled data set.  In addition, it serves as an examplar object for identifying procedures that can benefit the timing of the less-scattered pulsars included in pulsar timing arrays.   

The analysis of PSR J1903+0327 employed detailed modeling of its intrinsic pulse shape vs. frequency. For each of a set of candidate PBFs, the best fit intrinsic shape was determined based on a fit to the average of the PSR J1903+0327 profile over time. The resulting shape $U$ was model fitted to measured shapes to find the best fit PBF using a $\chi^2$ criterion. Candidate  PBFs included those originating from thin screens and from fully extended media.   The fitting procedure led to a three component model for the intrinsic shape and identification of a 
best case PBF  from a thin screen with 
Kolmogorov-like fluctuations and a small inner scale. 

Using multiepoch data from NANOGrav's PTA program, we analyzed variations of
$\DM, \tau$, and $X_\tau$ over a 5.5\,yr time span.  We conclude that  variations in 
the scattering parameters  are consistent with a model where they are modulated by refraction.  

Refraction occurs from density fluctuations on scales much larger than those responsible for diffraction, yielding scattered images with variable shapes and centroids that wander about the  direct line of sight. For a thin screen,  delays from path length differences are directly related to the angles of  arrival.   These changes are seen directly in the time series for $\tau$ and $X_\tau$.    For an extended medium the results are similar but there is no direct relation between AOA and time delay.  Instead, refraction alters  propagation along the entire ray path rather than only in a thin screen.
Refractive scintillation and refractive effects in pulsar dynamic spectra are other well known manifestations of  refraction.
Though demonstrated in Figs.\,\ref{fig:ref1}-\ref{fig:ref2}, yet  unproven, it seems much more reasonable to attribute changes in $\tau$ and $X_\tau$ to the happenstance refraction from a particular realization of scattering screen  rather than to actual changes in the properties of much smaller scale fluctuations.    The latter would require differences in physical conditions (e.g., energy injection or  magnetic field strength and direction) on scales smaller than $\sim1$\,au, which is less plausible for most lines of sight.

Correction of arrival times for the chromatic dispersion and scattering terms is problematic. The complexities of scattering \TOA\ shifts and \DM\ correction revealed for PSR J1903+0327 exist at varying degrees for all pulsars, whether of strong, moderate, or weak scattering.   Determination of \DM\ at each epoch requires correction for the scattering \TOA\ shifts.   Estimation of the latter requires knowledge of $U$ and the PBF, both of which are chromatic. However, the PBF shape and its $1/e$ width $\tau$ are epoch dependent, most likely caused by  epoch dependent refraction.  One approach to scattering correction might use the scattering time $\tau(\nu)$ itself as a \TOA\ correction, but this  will cause misestimation of \DM\  and its use in correcting the \TOA\  for the dispersive term in the timing equation Eq.\,\ref{eq:dmandscat}.  Alternatively, 
an empirical approach would include the last term in Eq.\,\ref{eq:dmandscat} in the timing model and solve for $\ttd$ and $X_t$ as part of the overall timing analysis.  This too will leave residual, chromatic errors in timing residuals, though they will be small for low-DM pulsars.   A more thorough analysis of  chromatic leakage into timing residuals  is beyond the scope of this paper but will be treated in detail in future work.

To improve pulsar timing errors, we suggest that more accurate arrival times $t_\nu$ can be obtained by  modeling  intrinsic and PBF shapes more rigorously.  This requires a customized treatment for each pulsar that takes into account the specific radio frequencies used in a timing program. 

This work was supported by National Science Foundation (NSF) Physics Frontiers Center award Nos. 1430284 and 2020265. P.R.B.\ is supported by the Science and Technology Facilities Council, grant number ST/W000946/1.
P.R.B.\ is supported by the Science and Technology Facilities Council, grant number ST/W000946/1.
M.E.D.\ acknowledges support from the Naval Research Laboratory by NASA under contract S-15633Y.
T.D.\ and M.T.L.\ are supported by an NSF Astronomy and Astrophysics Grant (AAG) award number 2009468.
E.C.F.\ is supported by NASA under award number 80GSFC24M0006.
D.R.L.\ and M.A.M.\ are supported by NSF \#1458952.
M.A.M.\ is supported by NSF \#2009425.
The Dunlap Institute is funded by an endowment established by the David Dunlap family and the University of Toronto.
T.T.P.\ acknowledges support from the Extragalactic Astrophysics Research Group at E\"{o}tv\"{o}s Lor\'{a}nd University, funded by the E\"{o}tv\"{o}s Lor\'{a}nd Research Network (ELKH), which was used during the development of this research.
S.M.R.\ and I.H.S.\ are CIFAR Fellows.
Pulsar research at UBC is supported by an NSERC Discovery Grant and by CIFAR.
S.J.V.\ is supported by NSF award PHY-2011772.

\bibliography{j1903scatt.bib}

\end{document}